\documentclass[onecolumn]{aastex6}
\usepackage{amsmath}

\def\kms{km ${\rm s}^{-1}$}
\def\sn1{${\rm s}^{-1}$}
\def\cmd{${\rm cm}^{-3}$}

\def\myxi{ergs cm ${\rm s}^{-1}$}
\def\ergs{erg ${\rm s}^{-1}$}

\def\sp{\space}

\def\cly{{\sc cloudy}}

\def\ni{\noindent}

\def\rmxaa{Rev. Mexicana Astron. Astrofis.}%

\shorttitle{Radiative tables for accretion}
\shortauthors{Ram{\'i}rez-Velasquez {\it et al.}}

\begin{document}

\bibliographystyle{apjsty}

\title{{\sc impetus}: New Cloudy's radiative tables
for accretion onto a galaxy black hole}

\author{Jos{\'e}~M.~Ram{\'i}rez-Velasquez$^{1,2}$, Jaime Klapp$^{2,3}$,
Ruslan Gabbasov$^4$, Fidel Cruz$^5$, Leonardo Di G. Sigalotti$^{1,5}$}
\affil{$^1$Centro de F\'{\i}sica, Instituto Venezolano de Investigaciones Cient\'{\i}ficas
(IVIC), Apartado Postal 20632, Caracas 1020A, Venezuela} 
\affil{$^2${\sc abacus}-Centro de Matem\'aticas Aplicadas y C\'omputo de Alto Rendimiento,
Departamento de Matem\'aticas, Centro de Investigaci\'on y de Estudios Avanzados
(Cinvestav-IPN), Carretera M\'exico-Toluca km. 38.5, La Marquesa, 52740 Ocoyoacac, Estado
de M\'exico, Mexico}
\affil{$^3$Departamento de F\'{\i}sica, Instituto Nacional de Investigaciones Nucleares (ININ),
Carretera M\'exico-Toluca km. 36.5, La Marquesa, 52750 Ocoyoacac, Estado de M\'exico, Mexico}
\affil{$^4$Instituto de Ciencias B\'asicas e Ingenier\'{\i}as, Universidad Aut\'onoma del Estado 
de Hidalgo (UAEH), Ciudad Universitaria, Carretera Pachuca-Tulancingo km. 4.5 S/N, Colonia
Carboneras, Mineral de la Reforma, C.P. 42184, Hidalgo, Mexico}
\affil{$^5$\'Area de F\'{\i}sica de Procesos Irreversibles, Departamento de Ciencias B\'asicas,
Universidad Aut\'onoma Metropolitana-Azcapotzalco (UAM-A), Av. San Pablo 180, 02200
Mexico City, Mexico}


\begin{abstract}

We present digital tables for the radiative terms that appear in the
energy and momentum equations used to simulate the accretion onto supermassive
black holes (SMBHs) in the center of galaxies.
Cooling and heating rates and radiative accelerations are calculated with two
different Spectral Energy Distributions (SEDs). One SED is composed of an
{\tt accretion disk + [X-ray]-powerlaw}, while the other is made of an
{\tt accretion disk + [Corona]-bremsstrahlung} with $T_X=1.16 \times 10^8$ K, where
precomputed conditions of adiabatic expansion are included. Quantification of 
different physical mechanisms at operation are presented, showing discrepancies
and similarities between both SEDs in different ranges of fundamental physical
parameters (i.e., ionization parameter, density, and temperature).
With the recent discovery of outflows originating at sub-parsec scales, these
tables may provide a useful tool to model gas accretion processes onto a SMBH.
\end{abstract}


\section{Introduction} 

Most of our present knowledge of the cosmos has come from application
of the principles of quantum mechanics and atomic physics. For instance,
the evolution of spectroscopy in every band of the electromagnetic
spectrum from radio to $\gamma$-rays has allowed the study of the supernova
remnants, the solar winds, the accretion onto supermassive black holes (SMBHs),
and the large-scale structure of the Universe in a way that we have never
imagined to be possible a century ago. Astrophysical processes that involve
radiative energy transfer are calculated by the balance between heating and
cooling. Analytical prescriptions for the heating and cooling rates in complex
environments are only possible under certain limits. 
Moreover, it is well-known that they depend on the SED used \citep{kallman1982a} 
and that stability curves also show a dependence on the SED in active galactic nuclei
\citep[eg.,][]{chakravorty2009a,chakravorty2012a}.
However, the increasing
computer power available today has allowed to model complex astrophysical
scenarios efficiently and at a relatively low cost, including the dynamical 
update of the microphysics and chemistry. Non-equilibrium thermodynamics,
ionization, molecular states, level populations, and kinetic temperatures of
low densities environments are some of the ingredients that have no analytical
counterparts and that can be calculated with highly efficient numerical
algorithms.

Among the several publicly available codes for the calculation of astrophysical
environments,
\cly \sp \citep{cloudy1303} 
and {\sc xstar} \citep{kallman2001a}
have become the most popular because they
treat the atomic physics at an
{\it ab-initio} level. 
In addition, they have
the ability to correctly handle a wide
variety of scenarios, while predicting the spectrum of different gas geometries,
including the Ultraviolet (UV) and the Infrared (IR) as well as a broad range of
densities up to $\sim 10^{15}$ \cmd\sp and temperatures from
the cosmic microwave background (CMB) to $10^{10}$ K. The electronic structure
of atoms, the photoionization cross-sections, the recombination rates, and the
grains and molecules are also treated in great detail.

In particular, the modeling in \cly \sp includes:
i) photoionization/recombination, ii) collisional ionization/3-body recombination
to all levels, and iii) collisional and radiative processes between atomic levels
so that the plasma behaves correctly in the low density limit and converges
naturally to local thermodynamic equilibrium (LTE) either at high densities or when
exposed to ``quasi-real" blackbody radiation fields \citep{ferland1998a}. Moreover,
collisions, line trapping, continuum lowering, and absorption of photons by
continuum opacities are all included as very general processes \citep{rees1989a}.
Inner-shell processes are also considered, including the radiative one (i.e.,
line emission after the removal of an electron) \citep{ferland1998a}. On the
other hand, analytical formulas for the heating and cooling rates have been
widely used. For instance, previous work on accretion onto SMBHs in the center
of galaxies (active galactic nuclei, AGNs) by \cite{psk00}, \cite{pk2004}, \cite{proga2007a},
and \cite{barai2011a}
have made use of \cite{blondin1994a} analytical formulas for the heating and
cooling rates, which are limited to temperatures in the range $10^4\lesssim T \lesssim 10^8$
K and ionization parameters ($\xi=L/[n_H r^2]$) in the interval
$1\lesssim \log(\xi) \lesssim 5$. 

In this paper, we develop a methodology and present tabulated values that account for highly 
detailed photoionization calculations together with the underlying microphysics to
provide a platform for use in existing radiation hydrodynamics codes based either
on Smoothed Particle Hydrodynamics (SPH) or Eulerian methods. Using the Cinvestav-{\sc abacus} 
supercomputing facilities, we have run a very extensive
grid of photoionization models using the most up-to-date version of \cly~({\bf v} 13.03), 
which allows us to pre-visualize physical conditions for a wide range of distances,
from four Schwarzschild radii ($\approx 4 r_{\rm Sch}$) to $r \lesssim 34,000 r_{\rm Sch}$
($r_{\rm Sch}=\frac{2GM_{BH}}{c^2}$), densities ($10^{-2}\lesssim n_H \lesssim 10^9$ \cmd),
and temperatures ($10^2\lesssim T \lesssim 10^9$ K) around SMBHs in AGNs.

It is well-known that accretion processes onto compact objects may influence the nearby
ambient around SMBHs in the center of galaxies
\citep[e.g.,][]{salpeter1964a,fabian1999a,barai2008a,germain2009a}.
Together with the outflow phenomena, they are believed to play a major role in the
feedback process invoked by modern cosmological models (i.e., $\Lambda$-Cold Dark Matter)
to explain the possible relationship between the SMBH and the host galaxy
\citep[e.g.,][]{magorrian1998a,gebhardt2000a} as well as in the self-regulating growth of the
SMBH. The problem of accretion onto a SMBH can be studied via hydrodynamical
simulations \citep[e.g.,][]{ciotti2001a,li2007a,ostriker2010a,novak2011a}. In numerical
studies of galaxy formation, spatial resolution permits resolving scales from the kpc to
the pc, while subparsec scales are not resolved. This is why a prescribed sub-grid is
employed to solve this lack of resolution. With sufficiently high X-ray luminosities,
the falling material will have the correct opacity, developing outflows that originate
at sub-parsec scales. Therefore, the calculation of the present tables provides a tool
to solve the problem of accretion onto SMBHs in the center of galaxies at sub-parsec 
scales. In addition, two SEDs and three 
ways of breaking up the luminosity between the disk and the X-ray components are presented.
On average, these runs take about 200 minutes using $\approx 4000$ cores ($\approx 13.3$k  
CPU hours) of the Cinvestav-{\sc abacus} supercomputer.

There are several radiation hydrodynamics codes that invoke \cly \sp for spectral 
synthesis. These codes are used to simulate processes subject to strong
irradiation such as the formation and evolution of HII regions, photoevaporation of the 
circumstellar disks, and cosmological minihaloes. For example, \cite{salz2015a} combine
a SPH-based magnetohydrodynamics (MHD) code with \cly \sp for the simulation of the
photoevaporation of the hot-Jupiter atmospheres. Moreover, \cite{niederwanger2014a} and
\cite{ottl2014a} combine a finite-volume MHD code with \cly \sp to simulate planetary
nebulae.  

The paper is structured as follows: in Section \ref{radicool}, we describe the SEDs
used and how they break up between UV and X-ray components. Details of the comparison 
between photoionization calculations using our two SED bases
are also provided. The calculation of the radiative acceleration as included in the
momentum equations is described in Section \ref{radiaccel}, while Section \ref{mytables}
contains details of the structure of the tables along with the meaning, units, and
location in the Internet for public use. The discussion of the results and the
conclusions are given in Section \ref{diss1}. Two appendices are added for the
description of the Sakura \& Sunyaev disk model and the calculation of the
ionic fractions. The symbols appearing through the manuscript have the standard meaning:
$G\equiv$ Newtonian gravitational constant, $c\equiv$ speed of light, $m_e\equiv$
electron mass, $M_{BH}\equiv$ black hole mass, $h\equiv$ Planck's constant,
$\sigma_T\equiv$ Thompson scattering cross-section, and $T\equiv$ temperature.

\section{Radiative Cooling and Heating} \label{radicool} 
 
In numerical simulations of accretion onto a BH with either SPH 
\citep[e.g., ][]{katz1996a}
or standard Eulerian methods \citep[e.g.,][]{kp09}, it is common practice to 
add the net radiative heating (or cooling, depending on the sign used) rate,
$\rho \mathcal{L}(\rho,T)=\mathcal{H}-\mathcal{C}$, into the energy equation as
\begin{equation}
\rho \frac{d}{dt}\left(  \frac{e}{\rho} \right)=-p \nabla \cdot
\boldsymbol{\upsilon} + \rho \mathcal{L},
\end{equation}
where $p,\rho, e$, and $\boldsymbol{\upsilon}$ 
are the pressure, density, energy density, and velocity of the gas,
respectively.
The heating ($\mathcal{H}[\rho,T]$) and cooling ($\mathcal{C}[\rho,T]$) 
rates are computed using \cly~13.03 \citep{cloudy1303}. A detailed
account of the techniques and atomic data can be found in the (very extended)
documentation of the code, namely Hazy1, Hazy2, and Hazy3. Therefore, many of the
details will not be repeated here, but rather we shall focus on describing all
the input parameters and code commands in order to accurately reproduce our
results. In brief, the problem reduces to have an abstract non-thermal 
equilibrium multidimensional unit cell (cloudy cell), which is able to return
pre-computed physical conditions, that is,
$\mathcal{H}$, $\mathcal{C}$, 
${\boldsymbol{\mathrm{g}}^{\rm rad}_{\rm Cont}}$,
${\boldsymbol{\mathrm{g}}^{\rm rad}_{\rm Grav}}$,
${\boldsymbol{\mathrm{g}}^{\rm rad}_{\rm Elec}}$,
${\boldsymbol{\mathrm{g}}^{\rm rad}_{\rm Line}}$,
and
${\boldsymbol{\mathrm{g}}^{\rm rad}_{\rm Total}}$ for given values of the 
hydrogen number density $n_H$, temperature $T$, distance to the source $r$,
and incident angle $\theta$.

In order for \cly \sp to handle the radiative transfer module, we must specify
the geometry to be employed. In particular, we use the {\sc wind} geometry in which
line widths and escape probabilities are evaluated either in the Sobolev or
Large Velocity Gradient (LVG) approximation. In this way, the effective line 
optical depth is
\begin{equation}
\tau_{l,u}(r)=\alpha_{l,u}\min (r,\Delta
r)\left(n_l-n_u\frac{g_l}{g_u}\right)\left(\frac{u_{\rm th}}{\max
(u_{\rm th},u_{\mathrm{exp}})}\right),
\label{wind1}
\end{equation}
\ni where $u_{\rm th}$ and $u_{\rm exp}$ are the thermal and expansion velocities,
respectively. The radius is chosen to be the minimum between the thickness of the
gas slab, $\Delta r$, and its distance from the ionizing source, $r$, which defines 
an effective column density smaller than the total cloud column density when the
radius is large and the expansion velocity is small. The population of the lower
and upper levels are $n_l$ and $n_u$, while their statistical weights are $g_l$
and $g_u$, respectively. The atomic absorption cross-section of the transition is
$\alpha _{l,u}$ (cm$^2$). Here we set a micro-turbulence velocity $u_{\rm turb}=100$
\kms \sp and an initial expansion velocity of 100 \kms. The thin shell approximation is
also invoked.

\begin{figure}
\rotatebox{0}{\resizebox{13cm}{!}
{\plotone{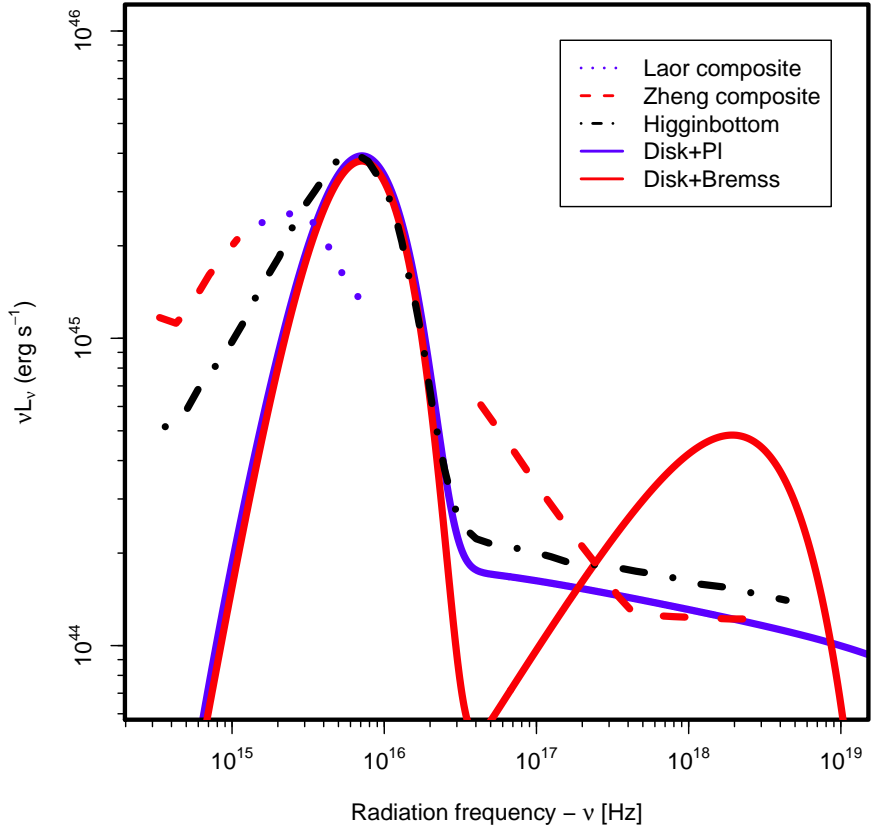}}}
\caption{
Multicomponent Spectral Energy Distributions (SEDs) used to calculate 
heating and cooling rates and radiative accelerations 
(SED1 (blue solid line): {\tt accretion disk + powerlaw} and SED2 (red
solid line): {\tt accretion disk + corona} with $ T_X=1.16 \times 10^8 $ K).
We have set them as close as possible to the SEDs used in \cite{pk2004}
and \cite{higginbottom2014a} (black dot-dashed line), which are in turn 
based on \cite{laor1997a} (blue dotted line) and \cite{zheng1997a} (red
dashed line) in order to facilitate comparison with previous work in the 
literature.
\label{SED1}}
\end{figure}

The input SEDs are shown in Fig. \ref{SED1}. The total luminosity is chosen to be
the typical luminosity of an AGN and is based on the accretion luminosity
$L_{\rm a}=2\eta G M_{BH} \dot{M}_a/r_{\rm Sch}$, where $\eta=0.0833$ is the accretion
efficiency and $\dot{M}_a=1.6$ $M_{\odot}$ yr$^{-1}$ \citep{proga2007a}. For a fiducial 
SMBH with $M_{BH}=10^8$ $M_{\odot}$, the total luminosity is set to
$L=L_{\rm a}=7.5\times 10^{45}$ \ergs. The two SEDs used are multicomponent spectra
similar to the ones observed for AGNs. The first one (i.e., SED1, blue solid line)
is composed of an {\tt accretion disk + [X-ray]-powerlaw}, while the second one (i.e.,
SED2, red solid line) is made of an {\tt accretion disk + [Corona]-bremsstrahlung}, with 
$T_X=1.16 \times 10^8$ K. The luminosity of the disk is defined as
$L_{\rm disk}=f_{\rm disk}L$, where $f_{\rm disk}=0.95$, 0.8, and 0.5 (see Table \ref{tbl1}),
while the luminosity of the X-ray power-law is $L_{\rm pl}=f_X ^{\rm pl} L$, with 
$f_X ^{\rm pl}=0.05$, 0.2, and 0.5. The energy index of the power-law is set to
$\Gamma_{\rm X}=2.1$ (with low- and high-energy cutoff equal to $10^5$ and $10^{10}$ K, 
respectively) to allow comparison with \cite{higginbottom2014a} (black dot-dashed line). 
The luminosity of the corona (in SED2) is set to $L_{X}=f_X^cL$, with $f_X ^c=0.05$, 0.2,
and 0.5. Table \ref{tbl1} gives a summary of the SED fractions used in the construction of the
tables, and we call them calculations {\sc i, ii, iii, iv, v}, and {\sc vi},
respectively.

A photo-ionizing background radiation from radio to X-rays
\citep{ostriker1983a,ikeuchi1986a,vedel1994a} and the cosmic microwave background 
(CMB) have been considered, where the CMB temperature, $T_{\rm CMB}=T_0(1+z)$ (K),
is taken to be $T_0=2.725 \pm 0.002$ K \citep{mather1999a,wilkinson1987a}.

\begin{figure}
\rotatebox{-90}{\resizebox{12cm}{!}
{\plotone{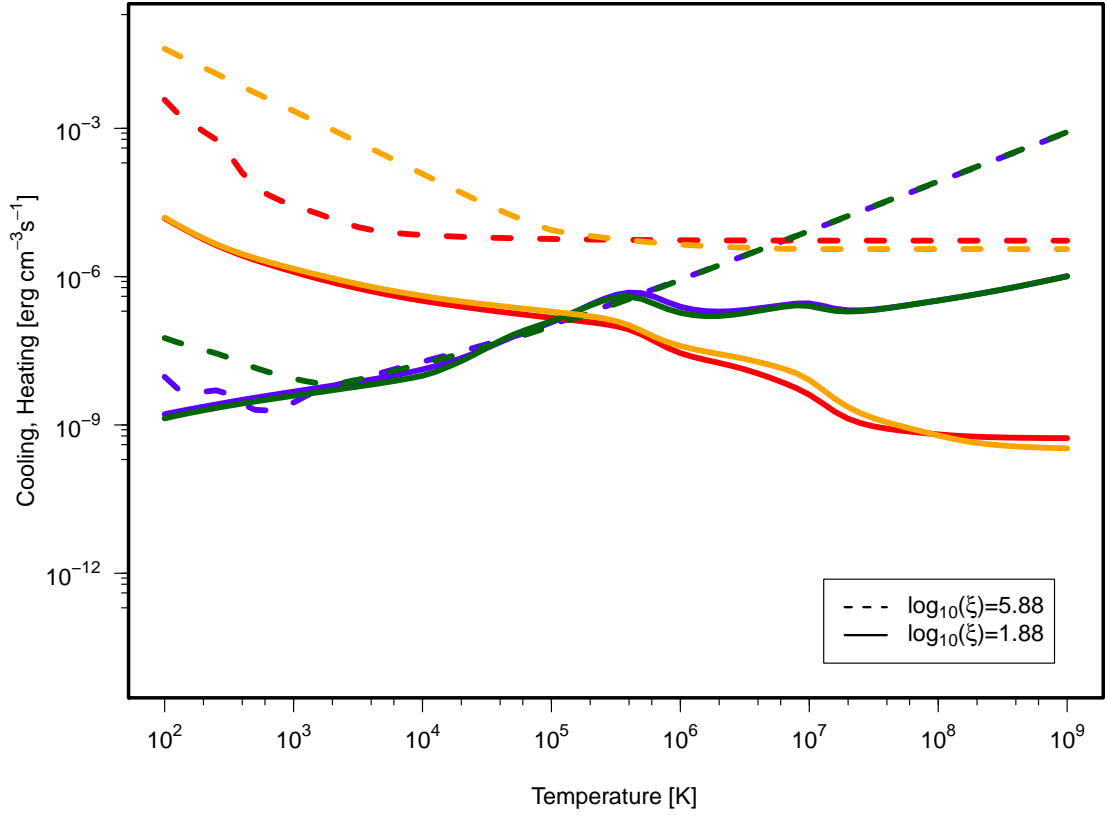}}}
\caption{
Cooling (blue and dark green lines) and heating (red and orange
lines) rates calculated with SED1 and SED2, respectively.
The solid and dashed lines correspond to
ionization parameters
$\log _{10}(\xi)=1.88$ and $5.88$ [\myxi]
($n_H=10^8$ \cmd),
respectively.
For these plots we have used $\theta=0^{\circ}$, $f_{\rm disk}=0.5$,
$f^{\rm pl}_{X}=0.5$, and $f^{c}_{X}=0.5$.
\label{cool1}}
\end{figure}

\begin{figure}
\rotatebox{-90}{\resizebox{18cm}{!}
{\plottwo{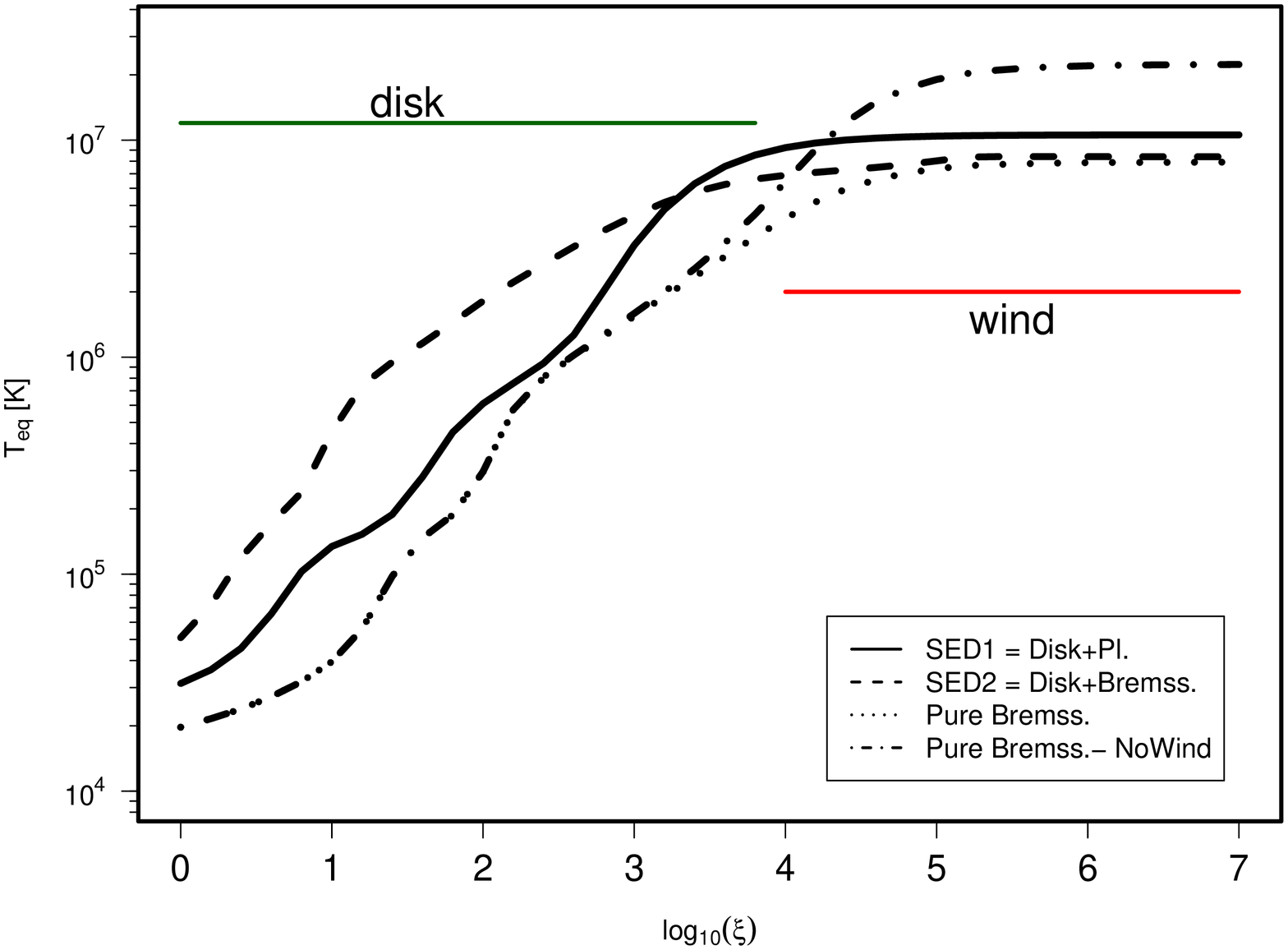}{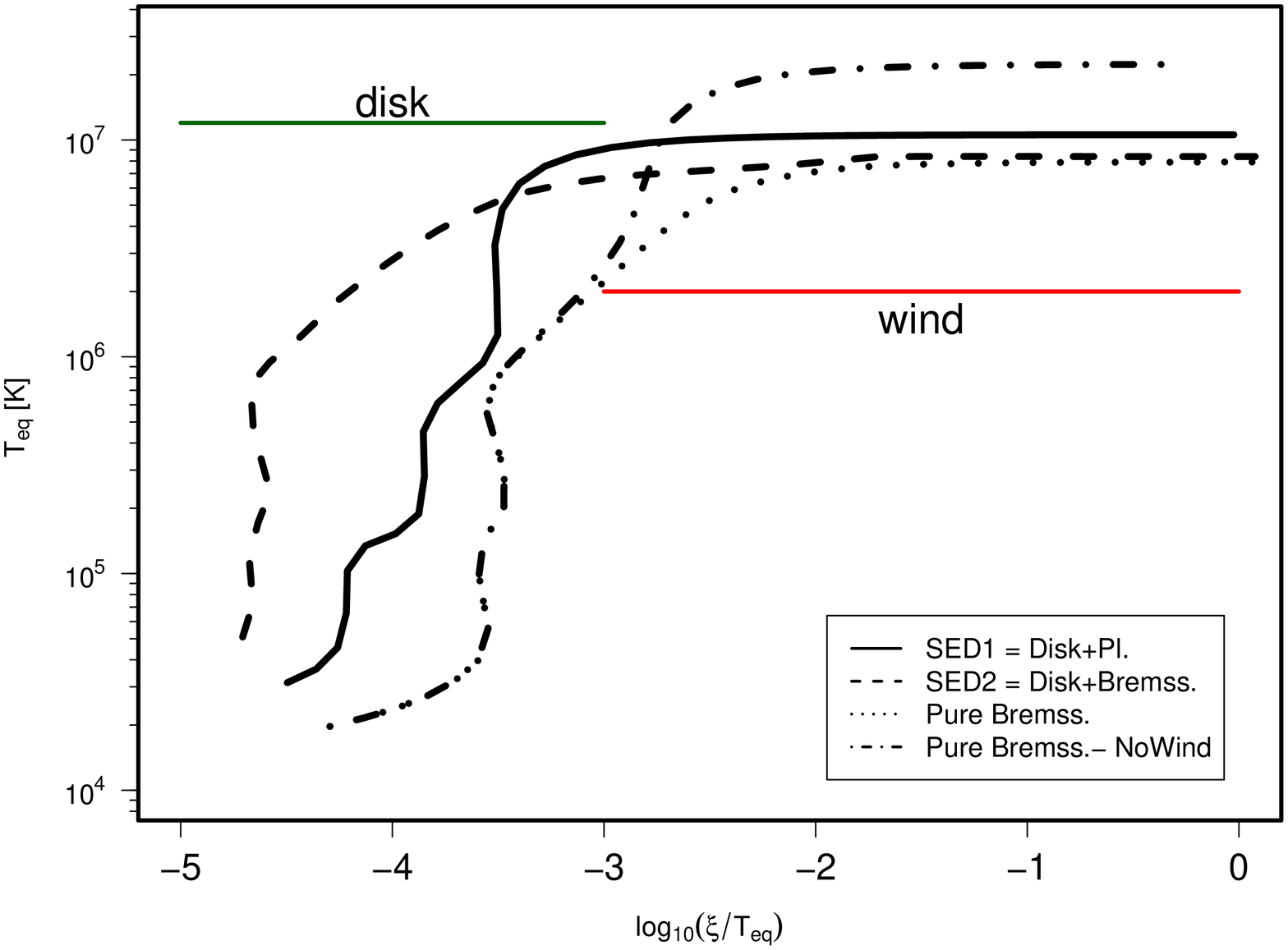}}}
\caption{
Gas equilibrium temperature as predicted by setting $\mathcal{H}=\mathcal{C}$
($\rho \mathcal{L}=0$) as a function of the ionization parameter
$\log_{10}(\xi)$ (top panel) and $\Xi = \log_{10}(\xi/T)$
(bottom panel), employed in models of accretion onto SMBH.
The solid line shows the predictions of
calculations {\sc i-iii} using a {\tt disk+X-ray powerlaw}.
The dashed line presents calculations of
{\tt disk+bremsstrahlung} (SED2 as a base).
The dotted and dashed-dotted lines correspond to calculations for a pure
10 keV bremsstrahlung SED with and without expansions effects.
See the text for details.
\label{Teq1}}
\end{figure}

A sample of the heating and cooling rates as determined from the tables is shown
in Fig. \ref{cool1}. The
gas has number density $10^8$ \cmd. In both
cases, $\mathcal{C}[\rho,T]$ and $\mathcal{H}[\rho,T]$ are calculated as functions
of the temperature for two different characteristic 
ionization parameters:
$\log _{10}(\xi)=5.88$ and $1.88$ [\myxi].
For comparison, the blue and red lines display the
total $\mathcal{C}[\rho,T]$ and $\mathcal{H}[\rho,T]$
for SED1.
The dark green and orange lines display the
total $\mathcal{C}[\rho,T]$ and $\mathcal{H}[\rho,T]$
for SED2.
In Fig. \ref{Teq1}, we depict the
gas equilibrium temperature predicted by $\mathcal{H}=\mathcal{C}$
($\rho \mathcal{L}=0$) as a function of the ionization parameter
$\log_{10}(\xi)$ for both the heating and cooling rates calculated
with the {\tt disk+corona} (e.g., calculations {\sc vi}, 
with $\theta\approx 2/5 \pi$) and
with the {\tt disk+X-ray powerlaw} presented in calculations {\sc iii}
(e.g., $\theta\approx \pi/2$, see Table \ref{tbl1}).
As stated by \cite{kallman1982a}, one of the main elements of the
photoionization calculations is the SED used.
For instance, we have over-plotted two more calculations:
one is a pure 10 keV bremsstrahlung (dotted line),
which includes; a) adiabatic cooling due to the hydrodynamic expansion
of the gas,
${\mathcal{C}_{\rm exp}} =  - \frac{p}{\rho }\frac{{D\rho
}}{{Dt}} + e\nabla  \cdot {\boldsymbol{\upsilon}},$
and b) Doppler shift due to the expansion.
The other is a pure 10 keV bremsstrahlung (dashed-dotted line),
but in this case we have relaxed the {\sc wind} model
(see Eq. \ref{wind1}, {\sc nowind} model) and
as a consequence adiabatic cooling and Doppler shift effects are not
taken into account.
As it can be seen the net effect is to drop the equilibrium temperature of
the gas from $\sim 2\times 10^7$ to  $\sim 7\times 10^6$ K in the range of photoionization
parameters $4 \lesssim \log _{10}(\xi) \lesssim 7 $.

Finally, we note an increase in $T_{\rm eq}$ in the range
$0 \lesssim \log _{10}(\xi) \lesssim 4 $, from the
{\sc nowind} bremsstrahlung model to the SED2 {\tt disk+bremsstrahlung},
which is basically explained by the presence of UV and hard photons
that are able to photoionize the gas and contribute with the
heating rate in that range of the ionization parameters.
The {\tt disk-blackbody}
component from the accretion disc affects the 
lower temperature part of the stability curve
\citep[see, for example,][]{chakravorty2012a}.
The {\sc nowind} bremsstrahlung model curve is similar
to the $T_{\rm rad}-\xi$ curve in Fig. 3 of \cite{barai2012a}.

In our calculations, the net Compton heating reads as follows
\begin{equation}
G_{\rm comp}-\Lambda_{\rm comp}=
\frac{4\pi n_e}{m_e c^2}\left[\int \sigma_h J_{\nu}\nu(1+\eta_{\nu})d\nu
-4kT\int\sigma_cJ_{\nu}d\nu \right],
\end{equation}
\ni where the Compton coefficients are $\sigma_h=\sigma_T\alpha$ and
$\sigma_c=\sigma_T\alpha\beta$, with
\begin{equation}
\alpha=[1+\nu_{Ryd}(1.1792\times 10^{-4}+7.084\times10^{-10}\nu_{Ryd})]^{-1},
\end{equation}
\ni and
\begin{equation}
\beta=[1-\alpha\nu_{Ryd}(1.1792\times10^{-4}+1.4168\times 10^{-9}\nu_{Ryd})/4],
\end{equation}
\ni where $\nu_{Ryd}$ is the photon frequency in Rydberg. The free-free heating rate
is defined as

\begin{equation}
G_{ff}=4\pi \int_{\nu_c}^{\infty} n_e \alpha_{\nu}(ff)J_{\nu}d\nu,
\end{equation}
\ni with $\alpha _{\nu}(ff)$ being the frequency-dependent free-free cross-section. 
The critical frequency
$\nu_c$ is defined as the frequency above which the gas at a depth $r$ into the cloud
becomes transparent. The free-free cooling rate is given by
\begin{equation}
\Lambda_{ff}=\int_{\nu_c}^{\infty} n_e \alpha_{\nu}(ff)4\pi B_{\nu}(T)d\nu,
\end{equation}
\ni where $B_{\nu}(T)$ is the frequency-dependent Planck function.

The net heating due to photoelectric and recombination cooling can be written as
\begin{equation}
G=G_{n,k}-\Lambda_{ind,n}-\Lambda_{spon,n},
\end{equation}
\ni where
\begin{equation}
G_{n,k}=n_n\int_{\nu_c}^{\infty} \frac{4 \pi J_{\nu}}{h\nu}\alpha_{\nu}
h(\nu-\nu_0)d\nu,
\end{equation}
\ni while the cooling rates due to induced recombination and spontaneous radiative 
recombination have the form
\begin{equation}
\Lambda_{ind,n}=n_e n_p P^*_n \int_{\nu_c}^{\infty} \frac{4\pi J_{\nu}}{h\nu}
\exp\left( -\frac{h\nu}{kT} \right)(\nu-\nu_0)d\nu,
\end{equation}
\ni and
\begin{equation}
\Lambda_{spon,n}=n_e n_p (\frac{2\pi m_e k}{h^2})^{-3/2}\frac{8\pi}{c^2}
\frac{g_n}{g_e g_{\rm ion}}T^{-3/2}\int_{h\nu_0}^{\infty} \nu^2 \alpha_{\nu}(n)
h(\nu-\nu_0)\exp\left[-\frac{h(\nu-\nu_0)}{kT}\right]d\nu,
\end{equation}
respectively.

Finally, the net heating due to collisional ionization and cooling by 3-body recombination is
defined as
\begin{equation}
G_{nk}-\Lambda_{nk}=\sum_nP^*_nn_en_pC_{n,k}h\nu_0(1-b_n),
\end{equation}
\ni where $C_{nk}$ is the collisional ionization rate, $P^*$ is the LTE
population, and $b_n$ is the departure coefficient.

The contribution from a line to the cooling rate is given by
\begin{equation}
\Lambda_{line}=h\nu_{ul}(n_lC_{lu}-n_uC_{ul}),
\end{equation}
\ni where $n_u$ and $n_l$ are the populations for the upper and lower levels,
respectively, and the $C_{ij}$ are the collision rates.
The SED1 calculated functions
depicted in Fig. \ref{cool1} show differences in shape and magnitude for
some temperature ranges compared to those obtained using SED2.

\begin{figure}
\rotatebox{0}{\resizebox{13cm}{!}
{\plotone{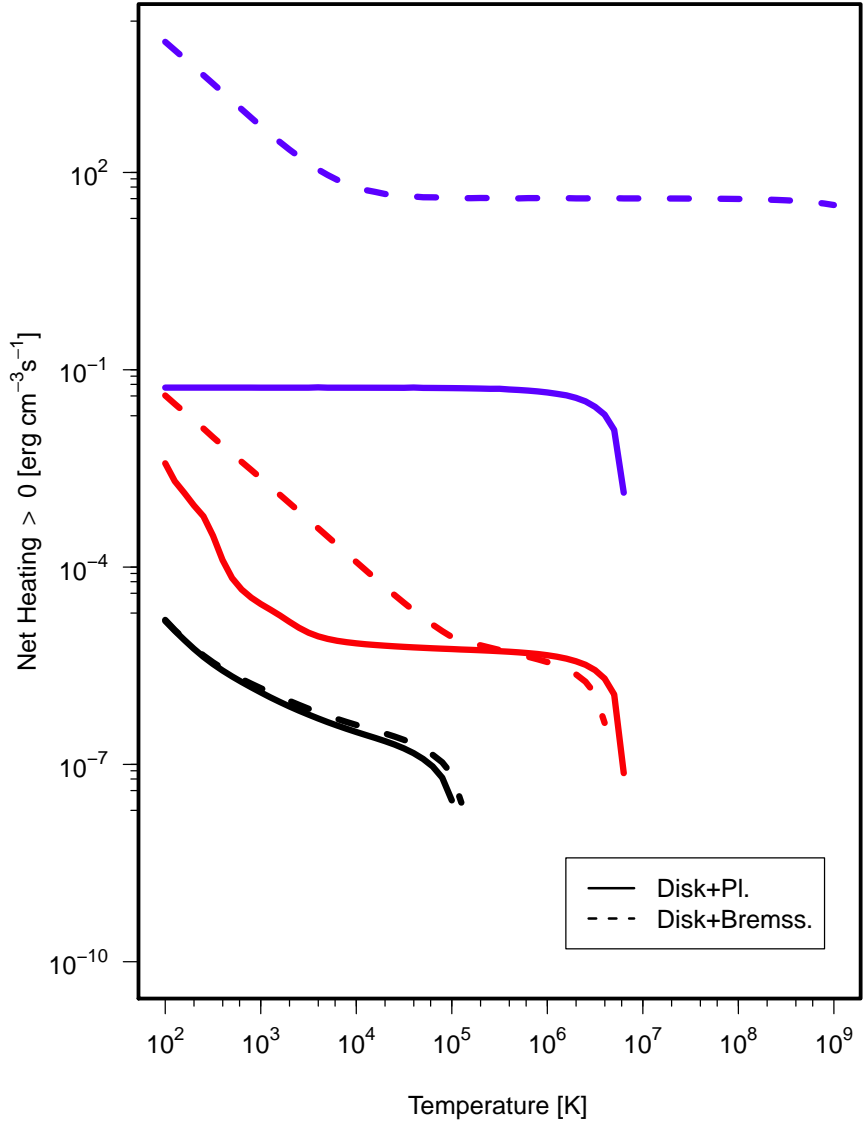}}}
\caption{
Net radiative heating $\rho \mathcal{L}$ as a function of temperature for the
three
different
ionization parameters
$\log _{10}(\xi)=9.88$ [\myxi] (blue lines),
$\log _{10}(\xi)=5.88$ [\myxi] (red lines), and
$\log _{10}(\xi)=1.88$ [\myxi] (black lines).
The dashed lines correspond to the SED2 calculations,
while the solid lines depict the net radiative heating from SED1.
In both sequences of curves $n_H=10^8$ \cmd,
$\theta=0^{\circ}$, $f_{\rm disk}=0.5$, $f^{\rm pl}_{X}=0.5$, and $f^{c}_{X}=0.5$.
\label{cool2}}
\end{figure}

Figure \ref{cool2} shows the net radiative heating 
($>~0$)
of the system for three characteristic
highly ionized plasmas with
$\log _{10}(\xi)=9.88,5.88$, and $1.88$ [\myxi]
(for $n_H=10^8$ \cmd)
usually found at
sub-parsec distances from the SMBH as obtained from SED2 calculations (dashed lines)
and SED1 calculations (solid lines). 
Important differences between both net radiative heatings
are observed for two of the distances tried, which may range from 50\% to
factors of a few percent
(as expected).
A higher level of the rate for all temperatures in the net heating 
is observed for calculations {\sc vi} at $\log _{10}(\xi)=9.88$, 
which leads to much warmer systems
at low temperatures compared to SED1's photoionization calculations. 
Farther away from the BH,
at ionization parameters
$\log _{10}(\xi)=5.88$
and in the temperature range
$10^2 \lesssim T \lesssim 10^4$
the heating exhibits a steeper slope due to the
removal of electrons from the pool by radiative recombination. 
This occurs because of
the large number of available transitions at these temperatures.
Nevertheless, at ionization parameters
$\log _{10}(\xi)=1.88$,
$\rho \mathcal{L}$ shows a very similar behavior.

\section{Radiative acceleration} \label{radiaccel}
 
Along with the heating and cooling rates, we also calculate the radiative acceleration,
which appears as a source term in the momentum equation
\begin{equation}
\rho \frac{d \boldsymbol{\upsilon}}{dt}=
-\nabla p + \rho {\bf g} + \rho {\boldsymbol{\mathrm{g}}^{\rm rad}_{\rm Total}}.
\end{equation}

In the tables we report here, the radiative acceleration, 
$\boldsymbol{\mathrm{g}}^{\rm rad}_{\rm Total}$ (defined as a force per unit mass), is 
calculated in a grid of $\theta$, $r$, $n_H$, and $T$ using \cly~13.03. For a direct 
attenuated continuum, $F_{\nu}$, and a density, $\rho$, we have that
\begin{equation}
{\boldsymbol{\mathrm{g}}^{\rm rad}_{\rm Total}} = \frac{1}{{\rho c}}\int {{F_\nu }{{\bar \kappa }_\nu }\;d\,\nu
+ } \frac{1}{{\rho c}}\sum\limits_l {{F_\nu }(l)\,{\kappa _l}\,{\gamma _l}}
\;{B_{l,u}}
\quad  [\mathrm{cm~s}^{-2}],
\label{grad1}
\end{equation}
where ${\bar \kappa }_\nu $ is the effective opacity from the continuum. The acceleration 
includes the usual photoelectric absorption as well as the free-free, Rayleigh, and Compton
processes. The integral is over the range between $\lambda \approx 10$~m and $h\nu=100$~MeV.
The second term is a summation over all lines contributing (typically $10^4$ to $10^5$
transitions). The quantity $\kappa _l$ is the opacity of the line, $B_{l,u}$ is the
Einstein coefficient, and $\gamma _l$ is the escape probability toward the ionizing source
(see Appendix B).

In Figs. \ref{rad1}(a)-(c), we display the variation of the total outward 
acceleration (solid lines), the radiative acceleration due to continuum processes
(dashed lines), and the acceleration due to spectral lines (dotted lines) with temperature
for three characteristic
ionization parameters:
$\log _{10}(\xi)=(9.88,5.88,1.88)$ [\myxi].
We may see from
Fig. \ref{rad1}(a) that 
in a highly ionized plasma with
$\log _{10}(\xi)=9.88$
(very close to the source, $\approx 4 r_{\rm Sch}$), the contribution is
mostly due to scattering. At 
$\log _{10}(\xi)=5.88$
($\approx 340 r_{\rm Sch}$), Fig. \ref{rad1}(b) shows that the
acceleration due to lines dominates up to $\approx 10^3$ K and then falls sharply at
higher temperatures due to the contribution of continuum processes, while at
$\log _{10}(\xi)=1.88$
($\approx 34,000 r_{\rm Sch}$), the acceleration due to spectral lines dominates over
the entire range of temperatures, except for $T \gtrsim 6\times 10^7$ K, as shown in 
Fig. \ref{rad1}(c). The force multiplier as a function of the temperature for the above three
characteristic values of $\xi$ is displayed in Fig. \ref{rad1}(d).

Moreover, in our calculations we assume that the contribution to
$\boldsymbol{\mathrm{g}}^{\rm rad}_{\rm Total}$, coming from the disk, depends on 
the radial direction ($r$) and the polar angle ($\theta$) through the incident radiation
\begin{equation}
F_{\rm disk}(r,\theta)=\cos(\theta)\frac{L_{\rm disk}}{4\pi r^2},
\end{equation}
\ni while the radiative flux from the central object is isotropic
\begin{equation}
F_{\rm co}(r)=\frac{L_{\rm co}}{4 \pi r^2}.
\end{equation}
In Fig. \ref{frad1}, we compare the radiative acceleration, 
$\boldsymbol{\mathrm{g}}^{\rm rad}_{\rm Total}$, as calculated from Eq. (\ref{grad1}) 
(black lines) for SED1, 
with $\boldsymbol{\mathrm{g}}^{\rm rad}_{\rm Total}$, given by SED2
(blue lines), both as
functions of the
ionization parameter
for three different angles, i.e., $\theta=$ $0^{\circ}$
(solid lines), $36^{\circ}$ (dashed lines), and $90^{\circ}$ (dotted lines).
When both approaches, i.e., the calculations {\sc i-iii}
and the calculations {\sc iv-vi}, are used in the range
$0 \lesssim  \log _{10}(\xi) \lesssim  15$
($10r_{\rm Sch}\lesssim r \lesssim 10,000r_{\rm Sch}$), 
they will both influence the distribution of velocities
resulting from the momentum equation in numerical simulations of the accretion onto
SMBHs.

\section{The Tables} \label{mytables}

The tables are available to the public at the following links:
{\sc {\tt www.abacus.cinvestav.mx/impetus} and 
\dataset[Zenodo]{https://zenodo.org/record/58984}}. 
They are plain ASCII files ({\tt my1Part\_OUT.txt})
stored in the directories {\tt simul\_$i$\_$j$/}, where the index $i$ corresponds 
to a value of the number density ($n_H$) and $j$ corresponds to a value of the incident
angle $\theta$. For example, the sub-directory {\tt simul\_$1$\_$1$/} contains the
ASCII text, with 12 columns (to be explained later), of the first density
($i\equiv 1$, $n_H=10^{-2}$ \cmd) and first angle ($j\equiv 1$, $\theta=0$) in
our grid. Moreover, in directory {\tt simul\_$101$\_$6$/}, one can find the calculations
for $n_H=10^{8}$ \cmd\sp and $\theta=\pi/2$.

Each main directory is provided with an ASCII file {\tt nH\_SED1\_mod\_11.txt},
where it is easy to see the values of $i$ and $j$ corresponding to a given density 
and angle. Inside this file we can find:

\begin{enumerate}

\item Column1: Index $i$.
\item Column2: Index $j$.
\item Column3: Number density $\log_{10}(n_H)$ [in \cmd].
\item Column4: Incident angle $\theta$ [in radians].
\item Column5: Initial radius $\log_{10}(r)$ [in cm].
\item Column6: Final radius $\log_{10}(r)$ [in cm].

\end{enumerate}

We now describe in more detail the content of the ASCII file 
{\tt my1Part\_OUT.txt}.
There are twelve (12) columns inside:

\begin{enumerate}

\item Column1: Incident angle $\theta$ [in radians].
\item Column2: Number density $\log_{10}(n_H)$ [in \cmd].
\item Column3: Distance from the BH $\log_{10}(r)$ [in cm].
\item Column4: Temperature $\log_{10}(T)$ [in K].
\item Column5: Total cooling rate $\log_{10}(\mathcal{C})$ [in erg \cmd s$^{-1}$].
\item Column6: Total heating rate $\log_{10}(\mathcal{H})$ [in erg \cmd s$^{-1}$].
\item Column7: Acceleration due to continuum
${\boldsymbol{\mathrm{g}}^{\rm rad}_{\rm Cont}}$ [in cm s$^{-2}$].
\item Column8: Acceleration due to gravity 
${\boldsymbol{\mathrm{g}}^{\rm rad}_{\rm Grav}}$ [in cm s$^{-2}$].
\item Column9: Total outward acceleration 
${\boldsymbol{\mathrm{g}}^{\rm rad}_{\rm Total}}$ [in cm s$^{-2}$].
\item Column10: Acceleration due to electron scattering 
${\boldsymbol{\mathrm{g}}^{\rm rad}_{\rm Elec}}$ [in cm s$^{-2}$].
\item Column11: Acceleration due to spectral lines 
${\boldsymbol{\mathrm{g}}^{\rm rad}_{\rm Line}}$ [in cm s$^{-2}$].
\item Column12: Force multiplier $M_{\rm t}$ [dimensionless].

\end{enumerate}

Two versions of the tables are made available: a short and a full version. The
short version contains only the {\tt my1Part\_OUT.txt} file, the illuminating SED at
$r=10^{14}$ cm (the {\tt my1contFile\_OUT\_14} file), and the ionic fractions at 
$r=10^{16}$ cm (the {\tt my1Part\_OUT\_frac.txt} file). On average, the short version
of the Tables (e.g. SED1, $f_{\rm disk}=0.95$ and $f_X=0.05$) occupies $\sim 47$ MB.
The full version contains the full output ({\tt my1Part\_OUT.out}) from \cly, which 
is useful to explore features related to the calculations in deeper detail. Each
uncompressed directory has, on average, a size of $\sim 32$ GB. Multiplying by six this size
leads to $\sim 200$ GB for the full version of the tables. A summary of the short and
full versions and their location can be found in Table \ref{tbl2}.
\section{Discussion and concluding remarks} \label{diss1}

The contribution of the microphysics to the heating and cooling rates is displayed
in Fig. \ref{heatAG1}. For instance, at $0.32$ pc we may see from Fig. \ref{heatAG1}(a) that
the main contributor to the heating over the temperature range $400\lesssim T \lesssim 8000$ K
is the Unresolved Transition Array (UTA, \cite{behar2001b,netzer2004a}
and also see \cite{ramirez2008b} for an observational point of view), 
which accounts for $\approx 12-16$\% of the
heating rate. In the 
interval $10^4\lesssim T \lesssim 8\times 10^4$ K, photoionization heating of O$^{+7}$
becomes the major contributor, providing from $\sim 12$ to 16\% of the heating. At
temperatures of $10^5\lesssim T \lesssim 3.2\times 10^5$ K, Fe$^{+18}$ contributes with
$\approx 12-17$\%. In these plots, the solid line labels the main contributors, while
the dotted and dashed lines depict the second and third contributors, respectively.
In Figs. \ref{heatAG1}(a)--(d), we see that in the temperature range 
$6.3\times 10^6\lesssim T \lesssim 3.2\times 10^9$ K, heating by Compton processes
dominate the heating with contributions that rise up to $\approx 100$\% close to the
upper extreme of the temperature range.

In general, we find a rather complex interplay between the different heating agents,
where low-ionization species contribute mostly at low-to-intermediate temperatures,
while highly ionized species of heavy metals (e.g., Fe$^{+17}$-Fe$^{+24}$) and 
intermediate heavy metals in the form of H- and He-like (e.g., O$^{+7}$-O$^{+8}$,
C$^{+4}$-C$^{+5}$) become important at temperatures in the range
$10^4\lesssim T \lesssim 10^6$ K. At $T\gtrsim 10^6$ K, Compton heating becomes
the dominant mechanism. The dashed-dotted lines in Figs. \ref{heatAG1}(a)--(d) 
depict the
contribution of the 100 main heating agents, which clearly account for $\approx 100$\%
of the total heating rate.

A similar analysis can be done for the cooling rate. In Fig. \ref{coolAG1}, we
depict the cooling rates as a function of the temperature at different distances
from the source. We see that radiative
recombination cooling by H contributes to $\approx 25-74$\% of the total cooling
rate in the temperature range $200\lesssim T \lesssim 4\times 10^4$ K. Cooling
by H lines dominates at higher temperatures in the interval
$5\times 10^4\lesssim T \lesssim 4\times 10^5$ K with a contribution to total
cooling of $\approx 25-74$\%. Free-free cooling contributes with up to $\approx 86$\%
in the temperature interval between $\sim 1.3\times 10^5$ and $\sim 1.3\times 10^8$ K,
with its contribution decreasing to $\approx 72$\% at $T\sim 10^9$ K. The second and
third contributors are represented by the dotted and dashed lines, respectively,
while the dashed-dotted lines depict the contribution of the 100 main cooling
agents.

Although the analytical formulas given by \cite{blondin1994a} are useful to study 
cooling and heating in high-mass X-ray binary systems, they become different
for simulations of gas accretion onto SMBHs in the center of galaxies,
if an accretion disk emission component and an expansion model
are adopted.
For
instance, \cite{barai2011a} discuss in detail three-dimensional SPH simulations of
accretion onto a SMBH, using the heating and cooling rates proposed by \cite{blondin1994a}
\citep[see also][for an Eulerian simulation]{mproga2013a}.
Some of their runs take longer to reach a steady state compared to the Bondi accretion.
When analyzing radiative properties in the $T-\xi$ plane, they find many particles
following the equilibrium temperature ($\mathcal{L}=0$) and discuss where and when
artificial viscosity plays a dominant role over radiative heating.

Below $T=10^4$ K, SED1 and SED2 cooling rates
differ by factors of a few.
In fact, neutral-to-middle ionized gas contributes
mostly to the total cooling below $T=10^4$ K and this can be important in the
outflows ($\sim 1400$ km/s) of {\sc n~iii}/{\sc n~iii}$^*$-{\sc s~iii}/{\sc s~iii}$^*$
found at $\sim$ 840~pc \citep{chamberlain2015a}, and also in closer Iron 
low-ionization broad absorption lines (FeLoBAL) flows ($v\sim$ few thousands \kms)
at $\sim 7-70$ pc \citep{mcgraw2015a}. Moreover, at $\log_{10}(\xi)\sim 2$
the SED2 computations may
overestimate the equilibrium temperature up to factors of $\sim 20$
in the range $10^5-10^8$ K.
This may be used as a discrimination feature for simulations of SEDs
in AGNs.
For the heating case, however, the differences only reach factors
of $\sim 2$ to $\sim 10$. We note that Compton and Coulomb heating could well be
operating at temperatures between $10^8$ and $10^{10}$ K 
\cite[for instance, in pre/post shocked winds in AGNs,][]{FaucherGiguere2012}.
In addition, pure 10 keV bremsstrahlung {\sc nowind} heating
and cooling differ less
from our calculations as the gas approaches the BH.

We further note that \cite{vignali2015a} found a gas of high velocity ($\sim 0.14$c)
through the identification of highly ionized species of Iron 
(e.g Fe~{\sc xxv} and Fe~{\sc xxvi})
in a luminous quasar at $z \sim 1.6$ located at distances of $\sim 10^{15}$--$10^{16}$ cm.
In fact, through observed
high-energy features, \cite{tombesi2015a} relate low- (by molecules) and high-velocity
(highly ionized gas) with the predicted energy conserved wind \citep{FaucherGiguere2012},
and locate this gas at $\sim$ 900 $r_{\rm Sch}$, where more precise estimates of the
heating and cooling are required. It is therefore clear that a quantitative analysis of
the heating and cooling agents operating on these kinds of astrophysical environments
are key aspects towards the understanding of the radiation hydrodynamical processes governing the
accretion onto SMBHs. We have provided the files {\tt my1Part\_OUT.het} and
{\tt my1Part\_OUT.col} as part of the tables, where the default $\approx 10$ agents 
are given by \cly. The interested reader may request the modified 100 agent files
to the corresponding author.

These tables have the potential and the flexibility to include other physical effects,
like dust and/or molecules. In fact, OH 119 $\mu m$ lines have been found in ultraluminous infrared galaxies
(ULIRGs)
using the Herschel/PACS telescope at velocities of $\sim 1000$ \kms \sp \citep{veilleux13a}.
Also, far-ultraviolet features may be present in Mrk 231 (found with the HST),
with velocities of $\sim 7000$ \kms \sp \citep{veilleux16a}.
They permit to make more extensive exploration about the
influence of the SED on photoionization calculations 
\citep[see][]{chakravorty2009a,chakravorty2012a}
and their impact on
the energy and velocity distribution on hydrodynamical accretion processes onto SMBH.
Another branch of SED to be explored are those including the reflected
spectrum from the accretion disk, a rich mix of radiative recombination continua,
absorption edges and fluorescent lines \citep{gar10,gar11,gar13a}.
Additionally, if produced close to the black hole, this component suffers
alterations due to relativistic effects \citep{dau13,gar14}.
These types of SEDs may influence cooling and heating rates
as they are very sensitive to the values of the
ionization parameter, temperature, and density.
In astrophysical ambients like the center of AGNs, they may play an important
role in high velocity winds and evolutionary stages of the host galaxies
\citep[as they may expel the cold gas reservoirs within $10^6-10^8$ years,][]{sturm11a}.
We also are in capacity to include them in tables of radiative acceleration for SPH codes,
and will be the subject of a future study.
A strict comparison between theoretical models and simulations is beyond the scope of
the tables presented here. At present, such simulations are under preparation.

\section{Acknowledgments}

{\sc impetus} is a collaboration project between the {\sc abacus}-Centro de
Matem\'aticas Aplicadas y C\'omputo de Alto Rendimiento of Cinvestav-IPN,
the Centro de F\'{\i}sica of the Instituto Venezolano de Investigaciones
Cient\'{\i}ficas (IVIC), and the \'Area de F\'{\i}sica de Procesos Irreversibles
of the Departamento de Ciencias B\'asicas of the Universidad Aut\'onoma 
Metropolitana--Azcapotzalco (UAM-A) aimed at the SPH modeling of
astrophysical flows. The project is supported by {\sc abacus} under grant
EDOMEX-2011-C01-165873, by IVIC under the project 2013000259, and by UAM-A
through internal funds. JMRV thanks the hospitality, support, and computing
facilities of {\sc abacus}, where this work was done.
We are also indebted to J. Garc\'{\i}a and M. Mel\'endez
for fruitful discussion.
We want to thank the anonymous referee for providing a number of
comments and suggestions that have improved both the content and
style of the manuscript.

\appendix 
\section{Geometrically thin, optically thick disk used in the SEDs}

The luminosity of a disk with dissipation $D(r)$ is
\begin{equation}
L_{\rm disk}=2\pi\int _{r_{iD}} ^{\infty} D(r) r dr = 
\frac{1}{2} \frac{\eta GM_{BH}\dot{M}}{r_{iD}},
\end{equation}
\ni which is half of the accretion luminosity $L_{\rm a}=\eta \dot{M}_{\rm a} c^2$.
If the disk is optically thick and its luminosity, $L_{\rm disk}$, radiates as a 
blackbody, its temperature $T_{bb}(r)$ as a function of distance is given by
\begin{equation}
\sigma _{\rm SB} T_{bb}^4=\frac{1}{2}D(r),
\end{equation}
\ni where $\sigma_{\rm SB}$ is the Stefan-Boltzmann constant and the factor
$\frac{1}{2}$ enters because only one side of the disk is considered. 
Using the form of $D(r)$ for a viscous accretion disk, we have that
\begin{equation}
T_{bb}(r)=T_{\rm iD}\left( \frac{r}{r_{\rm iD}} \right)^{-3/4}
\left[1-\left(  \frac{r_{\rm iD}}{r} \right)^{1/2}\right]^{1/4},
\end{equation}
\ni where
\begin{equation}
T_{\rm iD}=\left( \frac{3\eta GM_{BH}\dot{M}}{8\pi r_{iD}^3\sigma _{\rm SB}} \right)^{1/4}.
\end{equation}

For our SEDs we have used $M_{BH}=10^8~M_{\odot}$, $\dot{M}_{\rm a}=1.6~M_{\odot}$ yr$^{-1}$,
and $r_{\rm iD}=3r_{\rm Sch}$ ($\approx r_{\rm ISCO}$ for a non-rotating SMBH). Hence, in
the inner ring of the disk $T_{bb}(r_{\rm iD})=1.35\times 10^5$ K, while in the outer part,
i.e., for $r_{\rm oD}=10r_{\rm Sch}$, the temperature would be 
$T_{bb}(r_{\rm oD})=4.50\times 10^4$ K.


\section{\cly's Ionic fractions}

In our calculations we have included the following astrophysically relevant elements:
{H, He, C, N, O, Ne, Na, Mg, Al, Si, S, Ar, Ca, and Fe}. The abundances have been taken
from \cite{grass10} and we have neglected the effects of grains and molecules.
Our grid of \cly's models for the calculation of the heating and cooling rates and the
radiative acceleration uses the following physical parameters and resolutions:
\ni 
$\theta = 0...\pi/2$ with $\Delta \theta = \pi/10$;
$\log_{10} (n_H) = -2...9$ [\cmd] with $\Delta \log_{10} (n_H) = 0.1$;
$\log_{10} (r) = 14...18$ [cm] 
($\approx$ 3.4 $\times$ [$1-10^4$] $r_{\rm Sch}$)
with $\Delta \log_{10} (r) = 1$; and
$\log_{10} (T) = 2...9$ [K] with $\Delta \log_{10} (T) = 0.1$.

To look inside the cooling/heating tables we use a conventional bisection method,
where for each SPH particle (or Eulerian cell) with coordinates $(r_i,\theta _i, \phi _i)$
and density $\rho _i$, the functions $\mathcal{C}(\rho _i, T_i)$ and
$\mathcal{H}(\rho _i, T_i)$ are linearly interpolated within the temperature interval
$T_{\rm min} \leq T_i \leq T_{\rm max}$. \cite{cloudy1303} discuss in great detail the
numerical algorithm and the atomic databases used by \cly. Here we shall only describe the
calculation of the level populations and refer the interested reader to \cite{cloudy1303}
(and references therein) for technical details.

Radiative and collisional processes contribute to the evolution of the level populations
such that
\begin{equation}
\frac{db_n}{dt}=\left.\frac{db_n}{dt}\right\vert_{\rm rad} +
\left.\frac{db_n}{dt}\right\vert_{\rm col},
\label{Gbn}
\end{equation}
\ni where $b_n$ is the departure coefficient given by
\begin{equation}
b_n=\frac{n_n}{P^* _n n_e n_{\rm ion}},
\end{equation}
\ni $n_n$ is the actual population of the level, $n_e$ and $n_{\rm ion}$ are, respectively,
the electron and ion number density, and $P^* _n$ is the LTE relative population density for
level $n$ defined as
\begin{equation}
P^* _n=\frac{n^* _n}{P^* _n n_e n_{\rm ion}}=\frac{g_n}{g_e g_{\rm ion}}
\left( \frac{m^*_n}{m_{\rm ion}}  \frac{h^2}{2\pi m_e kT}\right)^{3/2}
\exp{\left(\frac{\chi _n}{kT}\right)}.
\end{equation}
\ni Here $g_n=2n^2$ is the hydrogenic statistical weight of level $n$, $n^*_n$ is the LTE
population of level $n$, $g_e=2$ is the electron statistical weight, $g_{\rm ion}$ is the ion
statistical weight, which is equal to 1 or 2 for H- or He-like species, respectively, and $\chi_n$ is
the ionization potential of level $n$. The other symbols are: the electron mass, $m_e$, the
Planck constant, $h$, and the temperature, $T$.

The collisional term in Eq. (\ref{Gbn}) can be written as
\begin{equation}
\begin{split}
\left.\frac{db_n}{dt}\right\vert_{\rm col}=
\sum_l b_l C_{nl} + \sum_u \frac{P^*_u}{P^*_n}b_u C_{un}
- b_n \left[  \sum_l C_{nl} + \sum_u \frac{P^*_u}{P^*_n} C_{un}
+ C_{nk} (1-b_n^{-1}) \right],
\end{split}
\end{equation}
\ni where the summations are taken over the upper and lower levels and the $C_{ij}$ are
the collisional rates in units of s$^{-1}$. The first, second, and third terms on the
right-hand side of the
above equation are, respectively, the collisional excitation from the lower levels to
level $n$, the collisional de-excitation to level $n$ from higher levels, and the term for
destruction processes. The collisional ionization rate, $C_{nk}$, is multiplied by a
factor that takes into account the effects of collisional ionization and three-body 
recombination.

The radiative contribution term in Eq. (\ref{Gbn}) can be written as
\begin{equation}
\begin{split}
\left.\frac{db_n}{dt}\right\vert_{\rm rad}=
\sum_l\frac{P^*_l}{P^*_n}b_lA_{nl}\frac{g_n}{g_l}\eta_{nl}\gamma_{nl}+
\sum_u\frac{P^*_u}{P^*_n}b_u(A_{un}\beta_{un}+A_{un}\eta_{un}\gamma_{un})+\\
\frac{\alpha_{\rm rad}+\alpha_{\rm ind}}{P^*_n}-b_n \times
\left[ \sum_l(A_{nl}\beta_{nl}+A_{nl}\eta_{nl}\gamma_{nl})+
\sum_uA_{un}\frac{g_u}{g_n}\eta_{un}\gamma_{un}+\Gamma_n \right],
\end{split}
\label{GbnRad}
\end{equation}
\ni where $A_{ij}$ is the transition probability,
$\eta_{ij}\equiv J_{\nu}(ij)/(2h\nu_{ij}^3/c^2)$ is the continuum occupation number of
the transition $ij$, with $J_{\nu}(ij)$ being the mean intensity of the ionizing
continuum at the line frequency $\nu$. The first of the two escape probabilities,
$\beta$, is a two-side function, which takes into account line scattering and escape
\begin{equation}
\beta(\tau,T)=\frac{\beta(\tau)+\beta(\cal{T}-\tau)}{2},
\end{equation}
\ni where $\tau$ is the optical depth of the point in question and $\cal{T}$ is the total 
optical depth. The escape probability, $\gamma_{ij}(\tau)$, accounts for the fraction of
the primary continuum penetrating up to $\tau$ and inducing transitions between level $i$ and
$j$.

The photoionization rate, $\Gamma_n$, from level $n$ that appears in Eq. (\ref{GbnRad}) is
given by
\begin{equation}
\Gamma_n=4\pi\int_{\nu_0}^{\infty} \frac{J_{\nu}}{h\nu}\sigma(\nu)d\nu,
\end{equation}
and the induced recombination rate (cm$^{3}$ s$^{-1}$) is defined as
\begin{equation}
\alpha_{{\rm ind},n}=P^*_n4\pi\int_{\nu_0}^{\infty} \frac{J_{\nu}}{h\nu}\sigma(\nu)
\exp{\left(-\frac{h\nu}{kT}  \right)} d\nu.
\end{equation}
Spontaneous radiative recombination rates, $\alpha_{\rm rad}$, are calculated as in
\cite{badnell2003a} and \cite{badnell2006a}.

In summary, we have added terms which correspond to induced upward transitions
from lower levels, spontaneous and induced downward transitions from higher
levels, spontaneous and induced capture from the continuum to the level, and
destruction of the level by radiative transitions and photoionization. The ionic
emission data is taken from CHIANTI \citep{dere1997a} and was recently revised by
\cite{landi2012a}. Figure \ref{iofrac1} shows the ionic fractions for all the
elements as a function of temperature. For these plots, we have chosen 
$n_H=10^{8}$ \cmd, $\theta = 0^{\circ}$, and a distance from the source equal
to $r=10^{16}$ cm (SED1, $f_{\rm disk}=0.8$, $f_X=0.2$).

\begin{deluxetable}{ccccc}
\tablecolumns{4}
\tablewidth{0pc}
\tablecaption{Emission fractions of the SEDs used for constructing the tables.  \label{tbl1}}
\tablehead{ 
\colhead{Calculation}&                 
\colhead{Base SED$^{\rm a}$}&   
\colhead{$f_{\rm disk}$}&   
\colhead{$f_{X}$}&       
\colhead{$T_{X}\times 10^6$ (K)$^{\rm b}$}
}
\startdata
{\sc i}       & 1 & 0.95    &   0.05    & 1.79 \\
{\sc ii}      & 1 & 0.8     &   0.2     & 7.00 \\
{\sc iii}     & 1 & 0.5     &   0.5     & 18.08 \\
{\sc iv}       & 2 & 0.95    &   0.05   & 1.11 \\
{\sc v}      & 2 & 0.8     &   0.2      & 4.39 \\
{\sc vi}     & 2 & 0.5     &   0.5      & 11.87 \\
\enddata
\tablecomments{
(a) Illuminating SED used as input for a given calculation: 
Base SED(1)$\equiv$ {\tt Disk + Pl} \citep{higginbottom2014a};
Base SED(2)$\equiv$ {\tt Disk + Bremss}.
(b) The Compton temperature depends on the ionization parameter
which changes with density and distance with a fixed luminosity.
This is $T_X$ for $n_H=10^8$ \cmd and $r=10^{16}$ cm. The entire
range of $T_X$ for the grid of parameters that we present in these tables
can be found in their large version (see Table \ref{tbl2}).
}
\end{deluxetable}

\clearpage

\begin{deluxetable}{ccc}
\tablecolumns{3}
\tablewidth{0pc}
\tablecaption{Table file versions and descriptions \label{tbl2}}
\tablehead{ 
\colhead{Calc}&
\colhead{File name}& 
\colhead{Size (MB)}
}
\startdata
{\sc i}       & \dataset[New\_DB\_SED1\_1\_short.tar.gz]{https://zenodo.org/record/58984} & 36 \\
{\sc ii}       & \dataset[New\_DB\_SED1\_2\_short.tar.gz]{https://zenodo.org/record/58984} & 47 \\
{\sc iii}       & \dataset[New\_DB\_SED1\_3\_short.tar.gz]{https://zenodo.org/record/58984} & 35 \\
{\sc iv}       & \dataset[New\_DB\_SED2\_1\_short.tar.gz]{https://zenodo.org/record/58984} & 49 \\
{\sc v}       & \dataset[New\_DB\_SED2\_2\_short.tar.gz]{https://zenodo.org/record/58984} & 49  \\
{\sc vi}       & \dataset[New\_DB\_SED2\_3\_short.tar.gz]{https://zenodo.org/record/58984} & 48 \\
\enddata
\tablecomments{
The main webpage of the project is: {\tt http://www.abacus.cinvestav.mx/impetus}.  The full version of the tables can be accessed by from the project webpage,
the links in the table above, or request to the corresponding author.
}
\end{deluxetable}



\begin{figure}
\rotatebox{0}{\resizebox{13cm}{!}
{\plotone{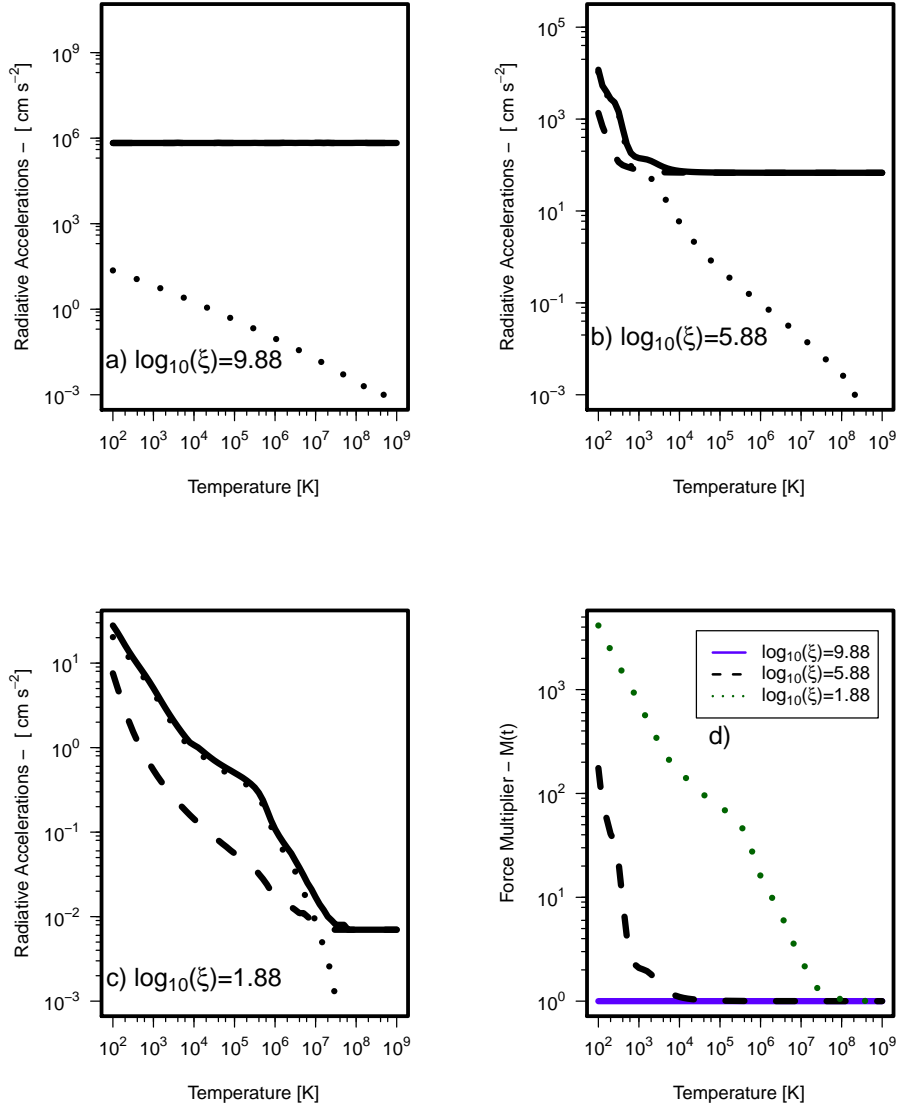}}}
\caption{
Calculated radiative accelerations for a BH of mass $M_{BH}=10^8 M_{\sun}$. Panels 
a), b), and c) show the total outward acceleration (solid lines), the acceleration
due to continuum processes (dashed lines), and that due to spectral lines (dotted
lines), respectively. Panel d) shows the force multiplier (M(t)$ \approx 1$: no
contribution from spectral lines and M(t)$ > 1$: line contribution different
from zero) for three characteristic
ionization parameters
$\log _{10}(\xi)=9.88,5.88$, and $1.88$ [\myxi].
In these plots we have used $\theta=0^{\circ}$, SED1, $f_{\rm disk}=0.5$,
$f^{\rm pl}_{X}=0.5$, and $n_H=10^8$ \cmd.
\label{rad1}}
\end{figure}
\begin{figure}
\rotatebox{-90}{\resizebox{17cm}{!}
{\plottwo{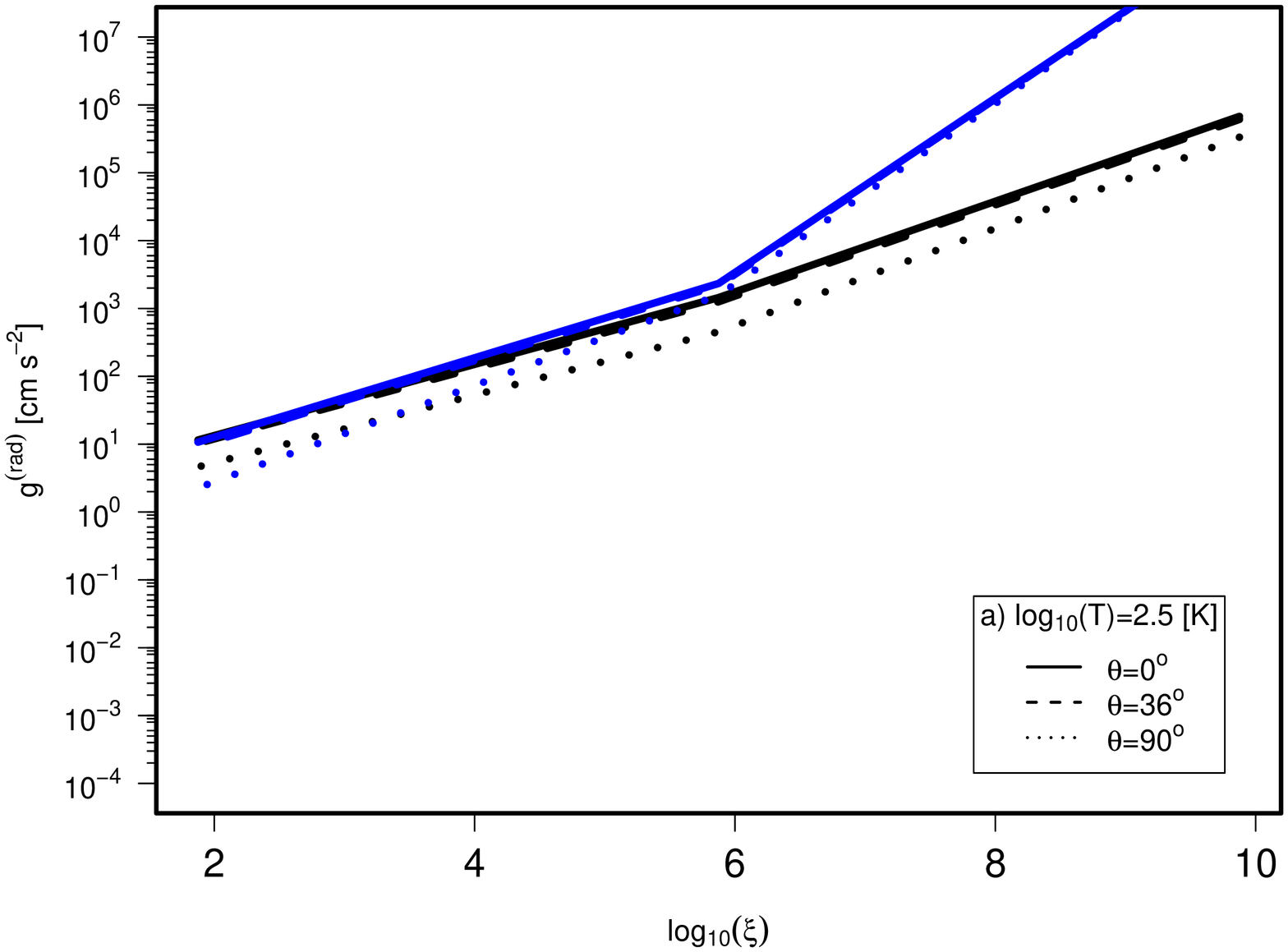}{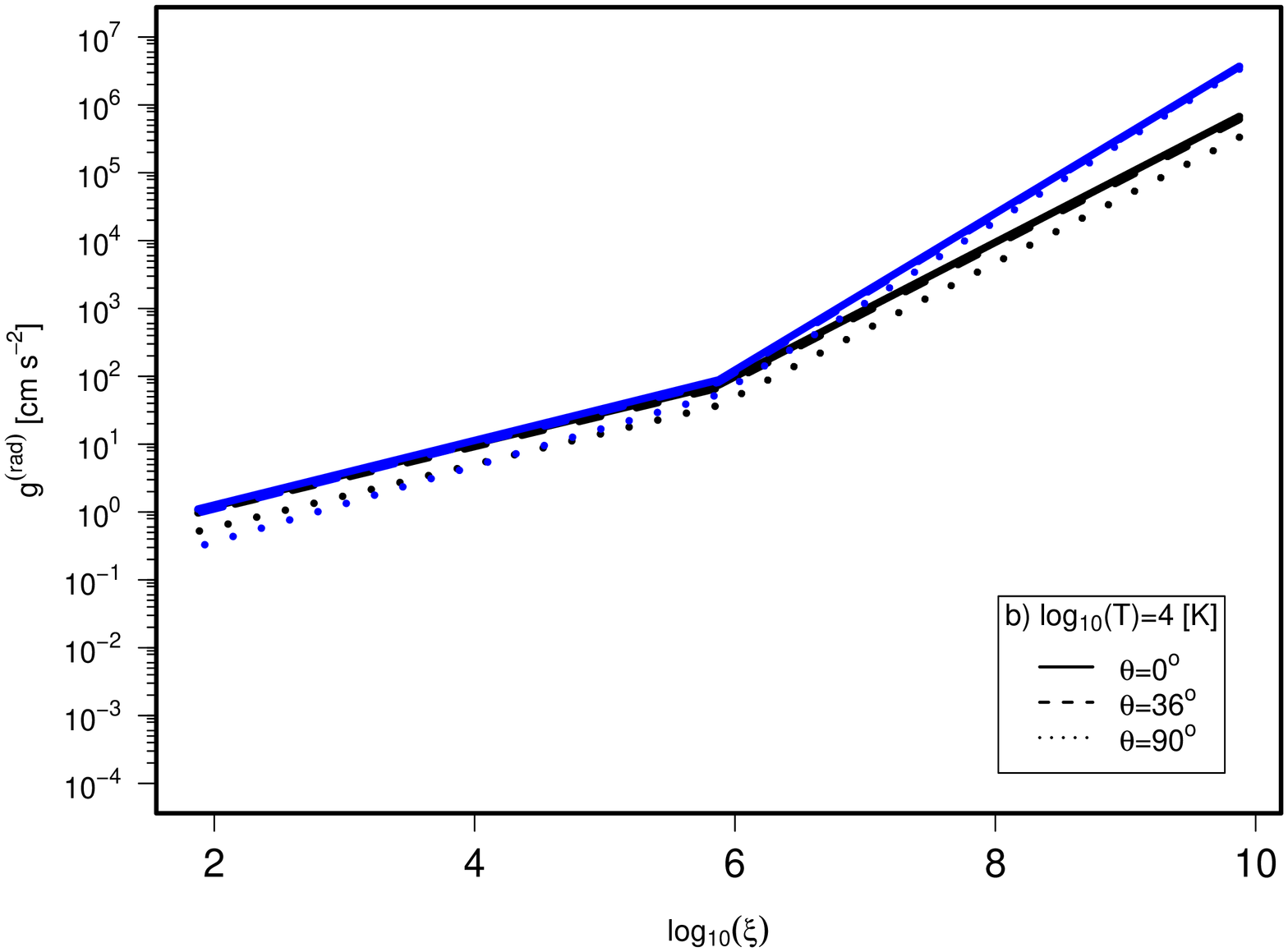}}}
\caption{
Total radiative force as it is usually included in the momentum equation for
the simulation of accretion onto a SMBH. The black lines depict the dependence
of the radiative force using calculation {\sc iii} (SED1)
on the ionization parameter $\xi$
for $\theta=0^{\circ}$ (solid lines),
$36^{\circ}$ (dashed lines), and $90^{\circ}$ (dotted lines) as obtained from
our calculations. The blue lines display the corresponding behavior as obtained
using calculation {\sc vi} (SED2)
radiative forces. In the upper panel (a), the gas temperature is fixed
to $\log_{10}T=2.5$ [K], while in the lower panel (b) $\log_{10}T=4$ [K]. When
$\log_{10}T=4$ [K], differences up to a factor of $\sim 10$ are seen close
to the source (distances are from $\sim 3,400r_{\rm Sch}$ to $34,000r_{\rm Sch}$).
\label{frad1}}
\end{figure}
\begin{figure}[h]
\begin{minipage}[b]{0.50\linewidth}
\centering
\includegraphics[width=0.90\linewidth]{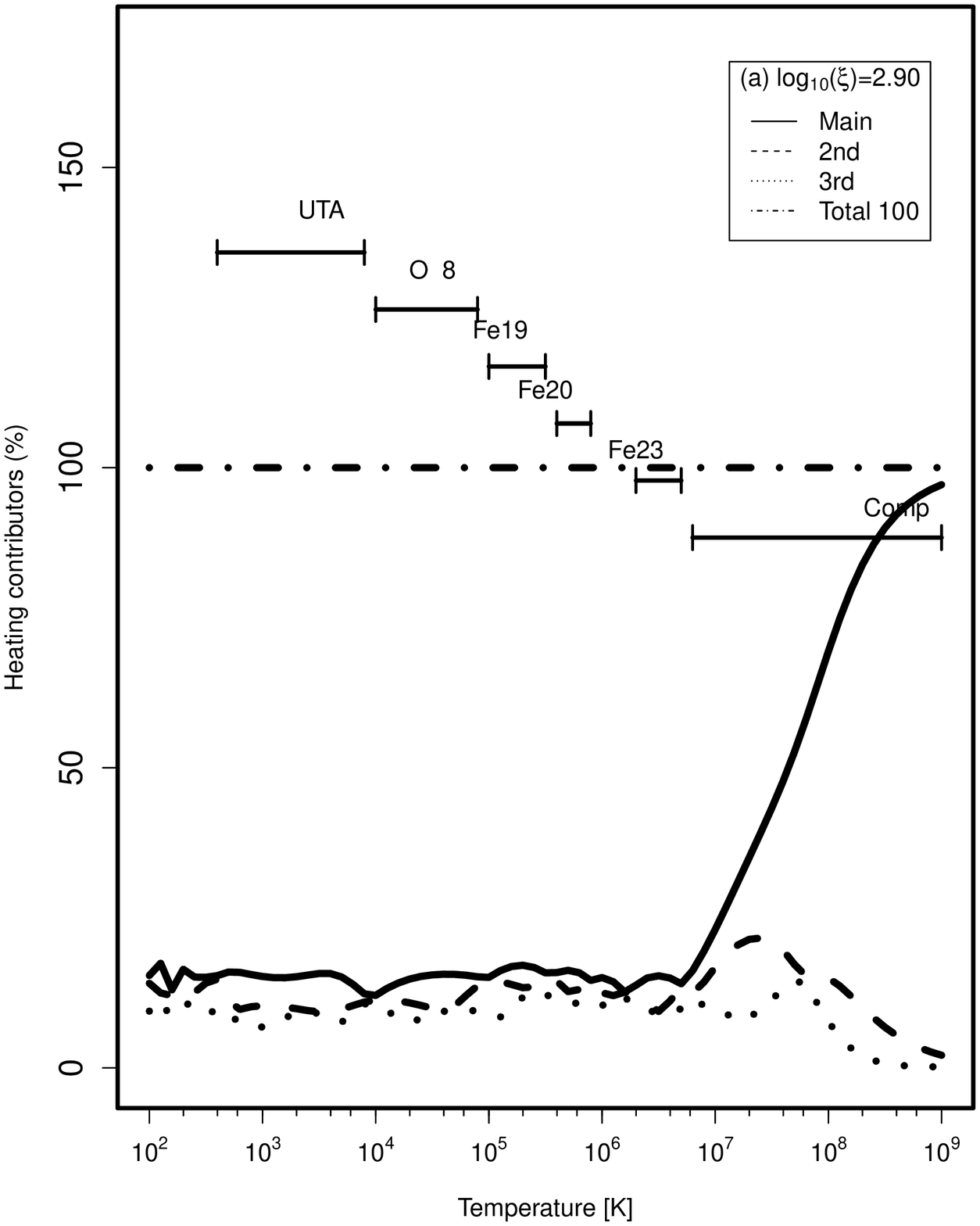}
\end{minipage} 
\begin{minipage}[b]{0.50\linewidth}
\centering
\includegraphics[width=0.90\linewidth]{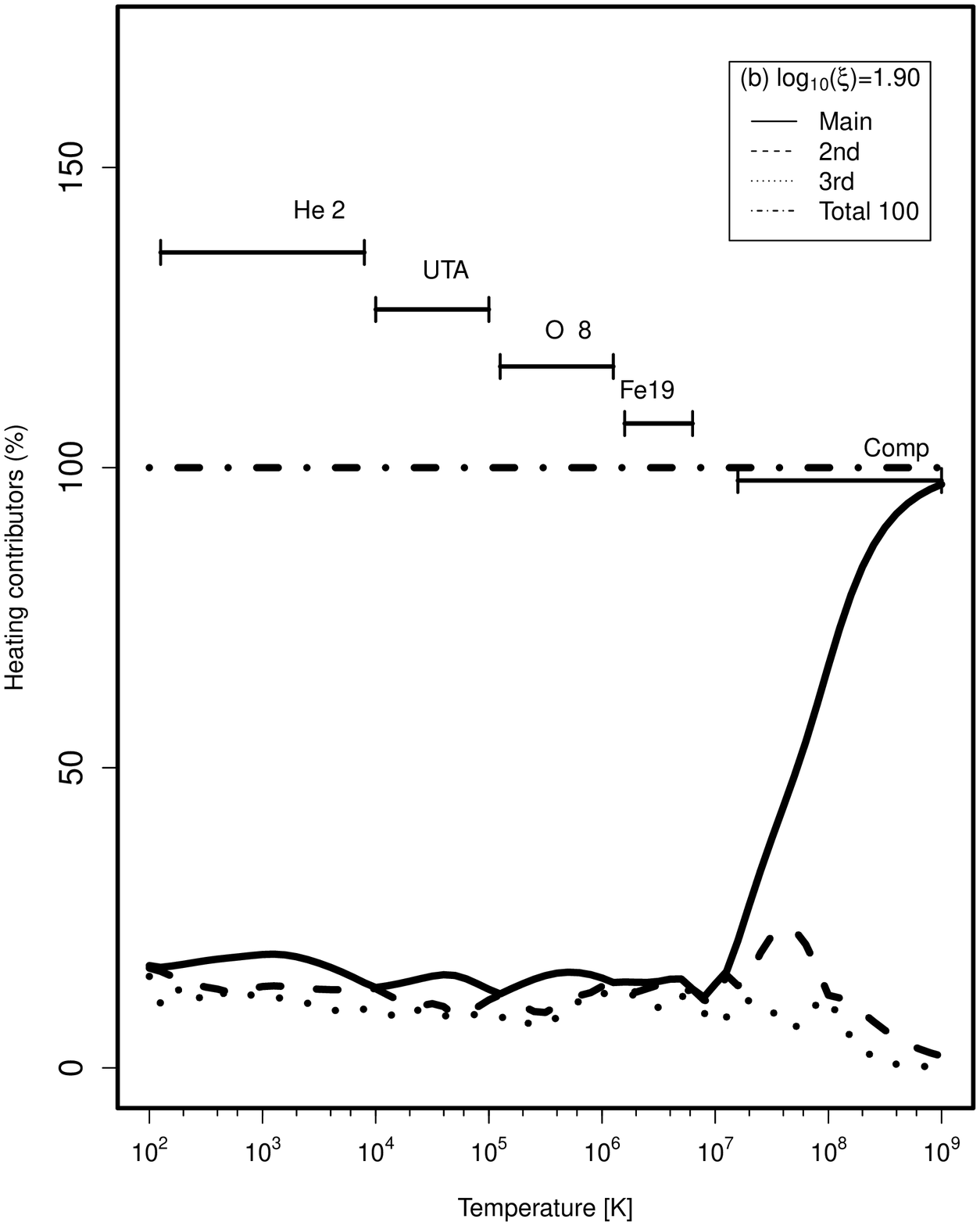} 
\end{minipage} 
\begin{minipage}[b]{0.50\linewidth}
\centering
\includegraphics[width=0.90\linewidth]{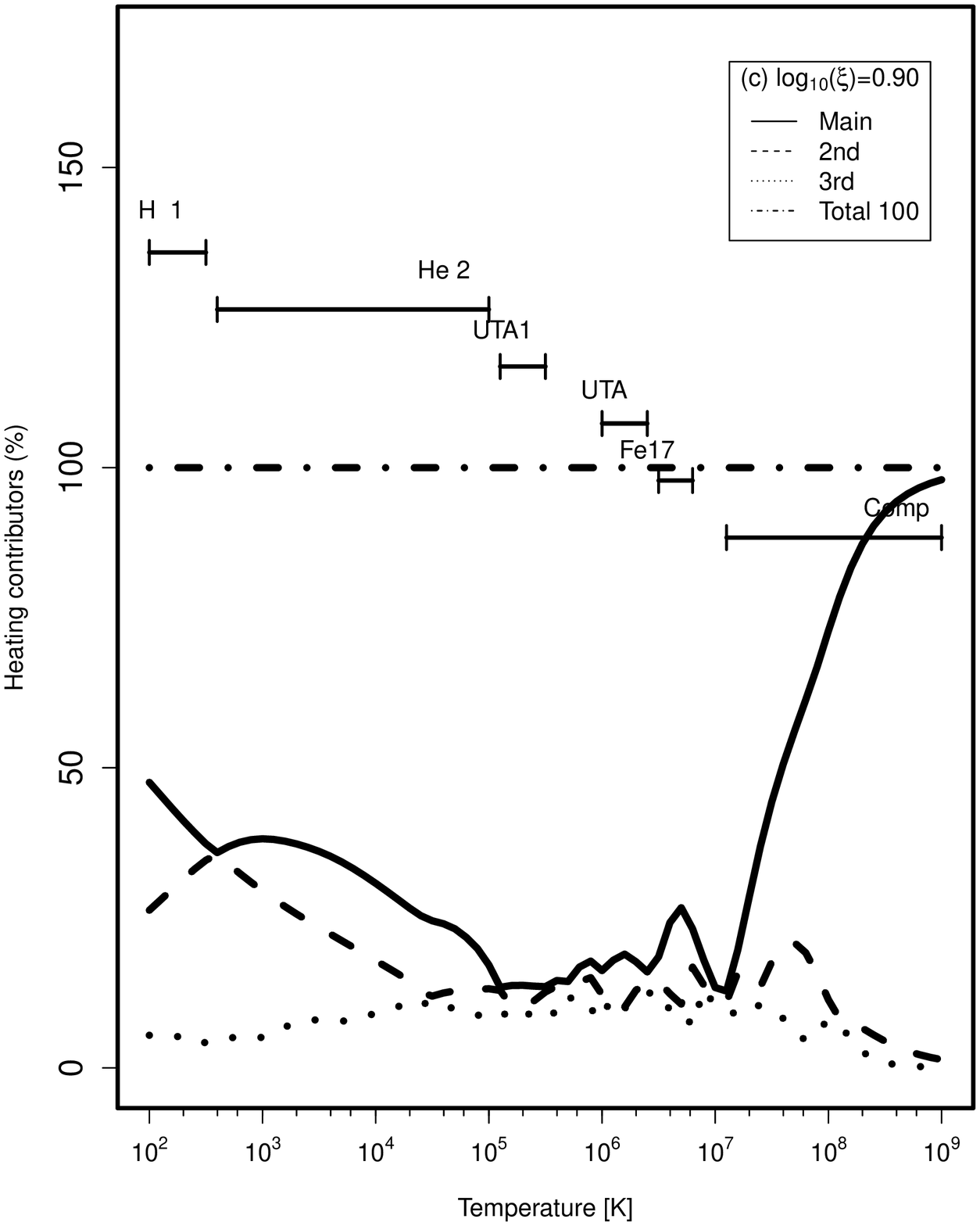}
\end{minipage} 
\begin{minipage}[b]{0.50\linewidth}
\centering
\includegraphics[width=0.90\linewidth]{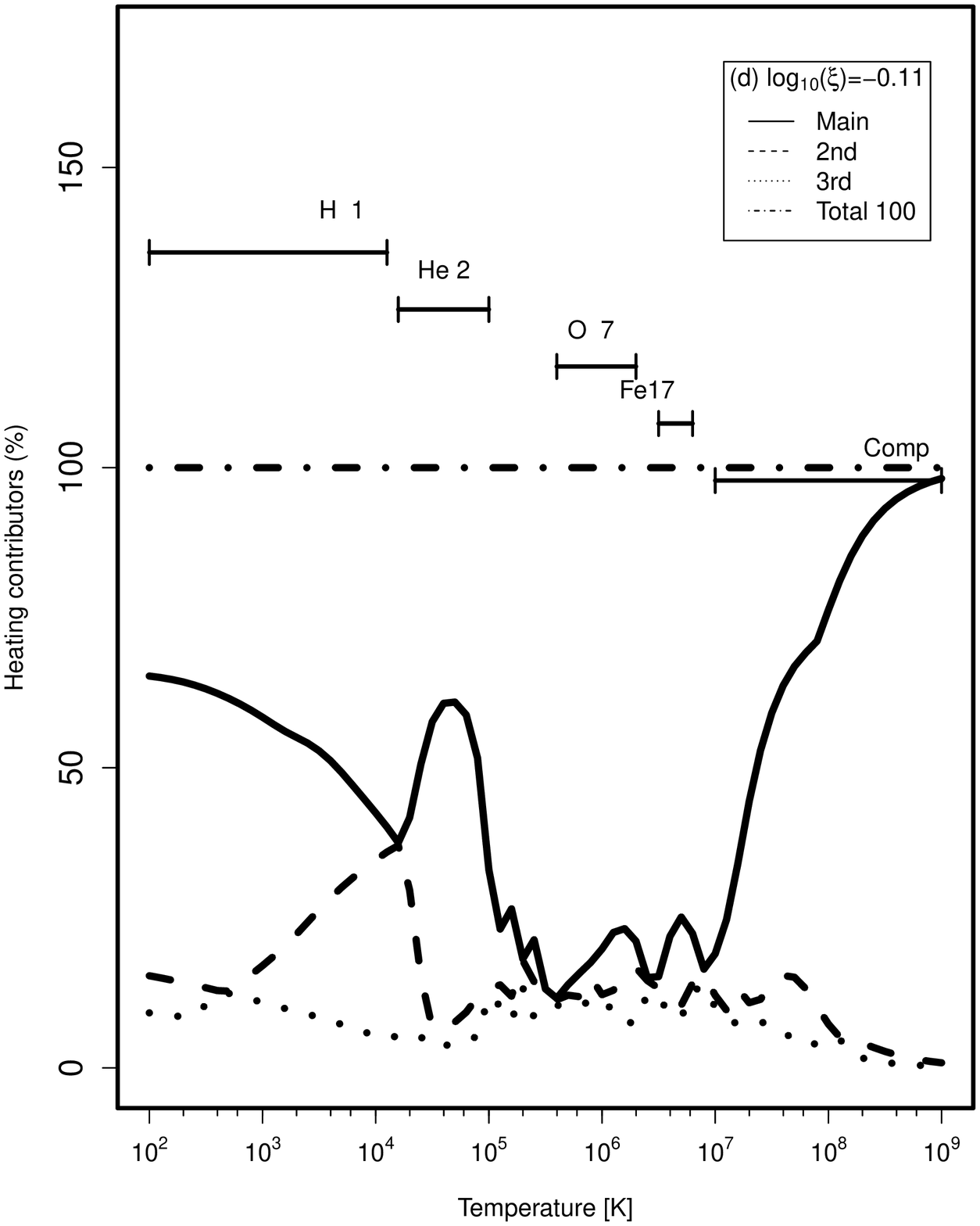}
\end{minipage} 
\caption{
Main agents contributing to the heating rates as included in the tables.
A complex interplay between photoelectric heating of low and high ionization species,
UTA, and Compton processes can be observed. For these plots we have used a
bremsstrahlung SED with $ T_X=1.16 \times 10^8$ K, in order to
directly see the different physical mechanisms at operation
for a gas with $n_H=10^7$ \cmd \sp at
ionization parameters $\log(\xi)$:
(a) 2.90, (b) 1.90, (c) 0.90, and (d) -0.11 [\myxi].
The main agents (solid lines)
are also labeled by straight horizontal lines to denote the temperature range
where they contribute. The second (dotted lines) and third (dashed lines)
contributors are also displayed. The dotted-dashed lines correspond to the
100 main heating agents.
\label{heatAG1}}
\end{figure}
\begin{figure}[h]
\begin{minipage}[b]{0.50\linewidth}
\centering
\includegraphics[width=0.90\linewidth]{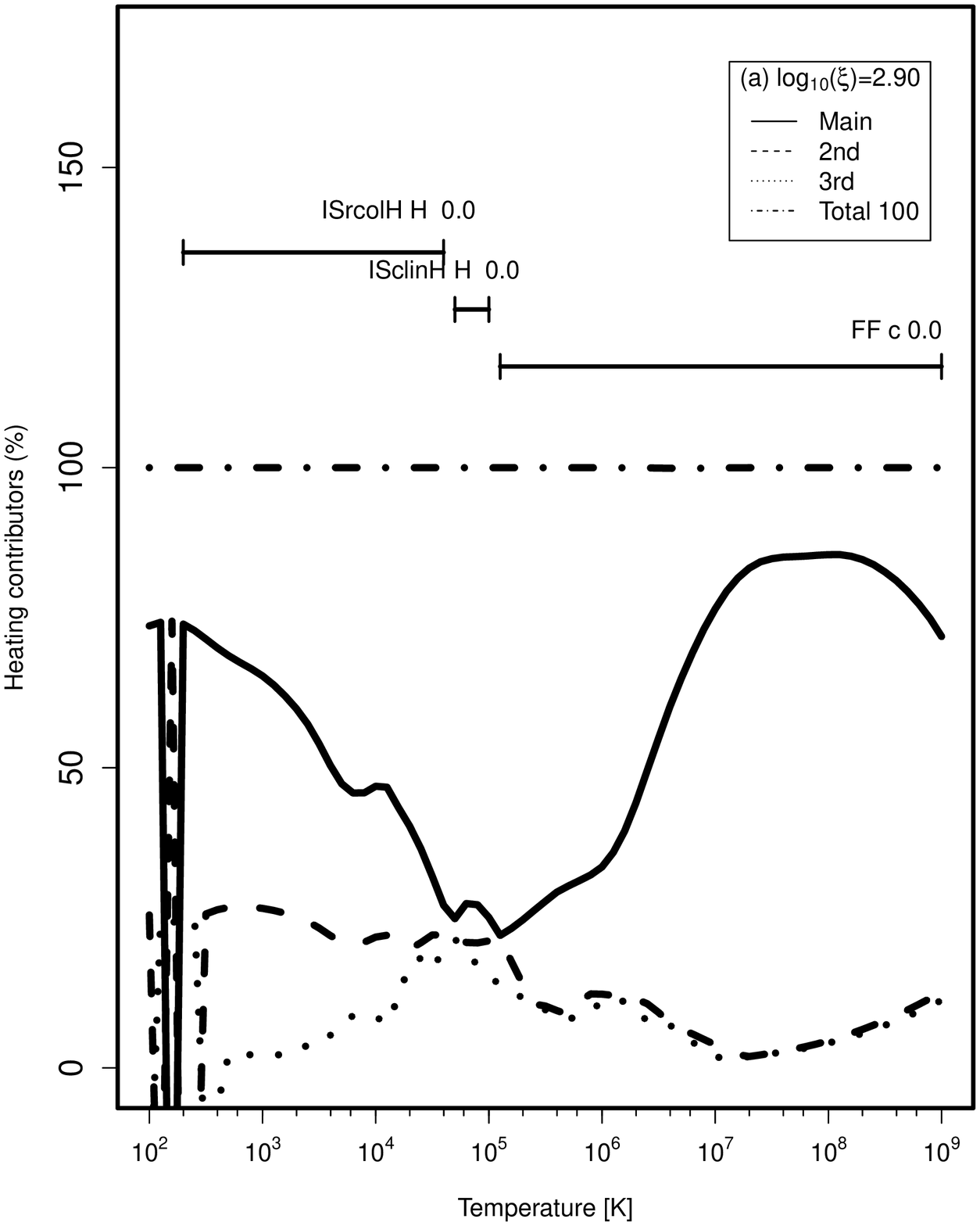}
\end{minipage} 
\begin{minipage}[b]{0.50\linewidth}
\centering
\includegraphics[width=0.90\linewidth]{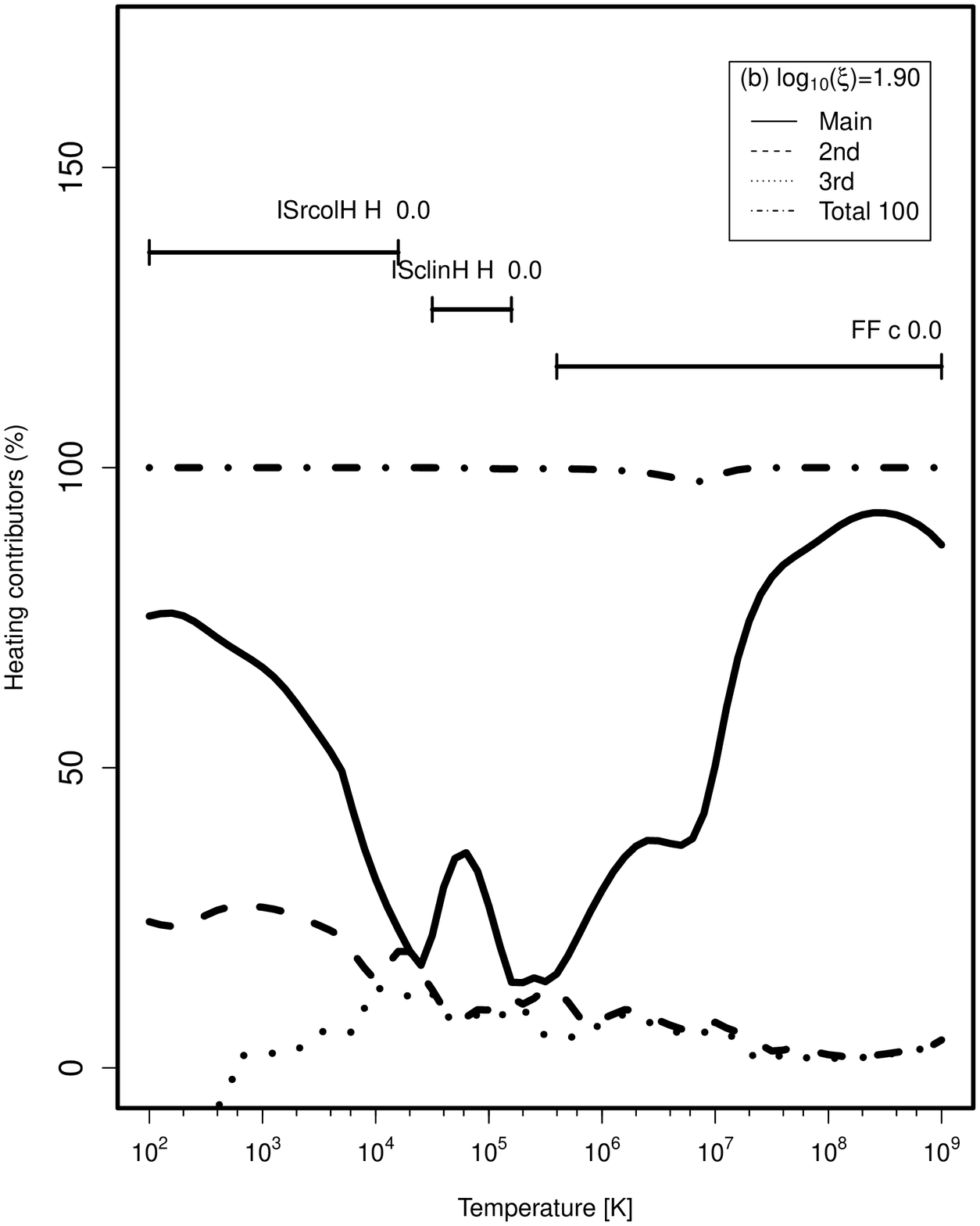} 
\end{minipage} 
\begin{minipage}[b]{0.50\linewidth}
\centering
\includegraphics[width=0.90\linewidth]{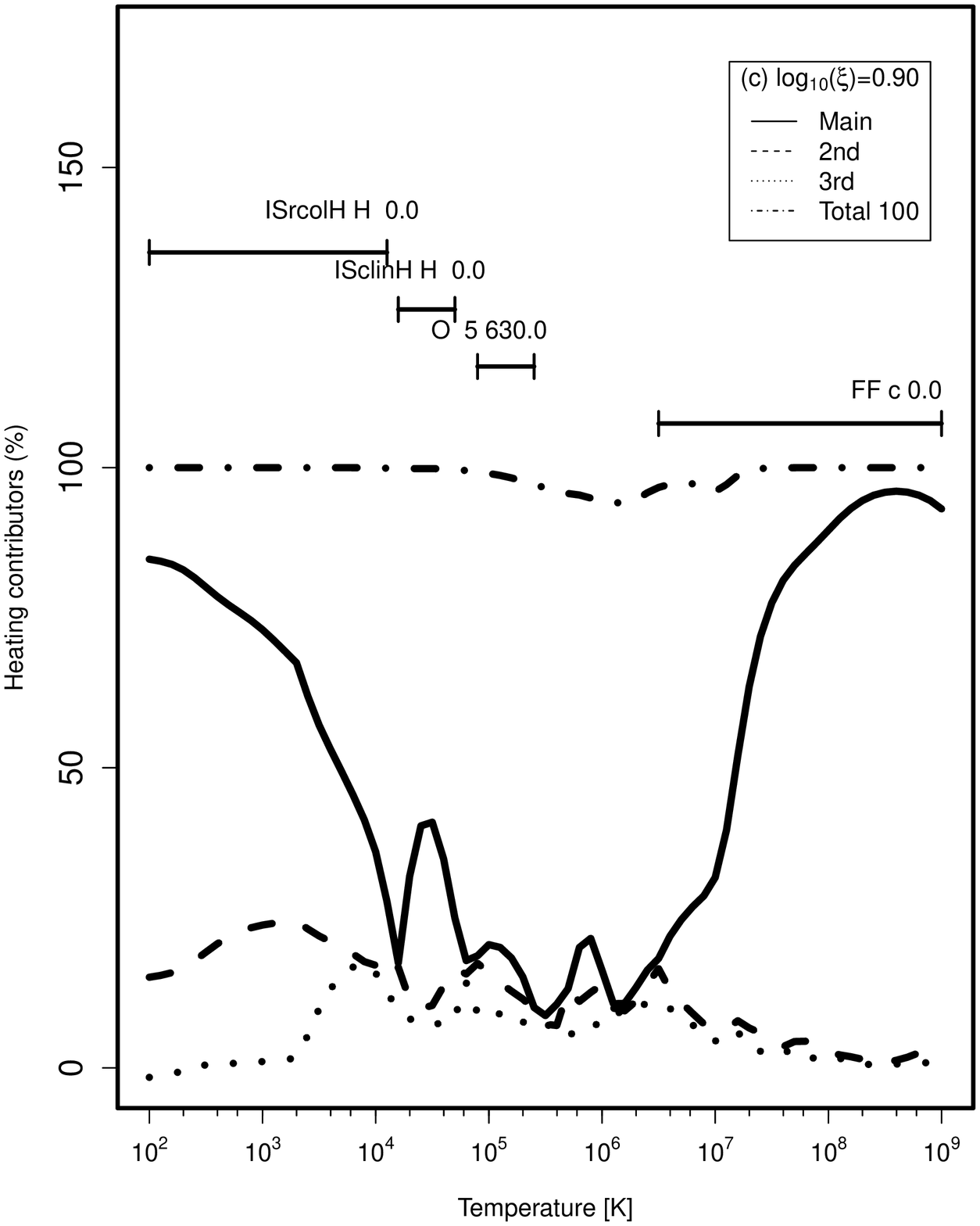}
\end{minipage} 
\begin{minipage}[b]{0.50\linewidth}
\centering
\includegraphics[width=0.90\linewidth]{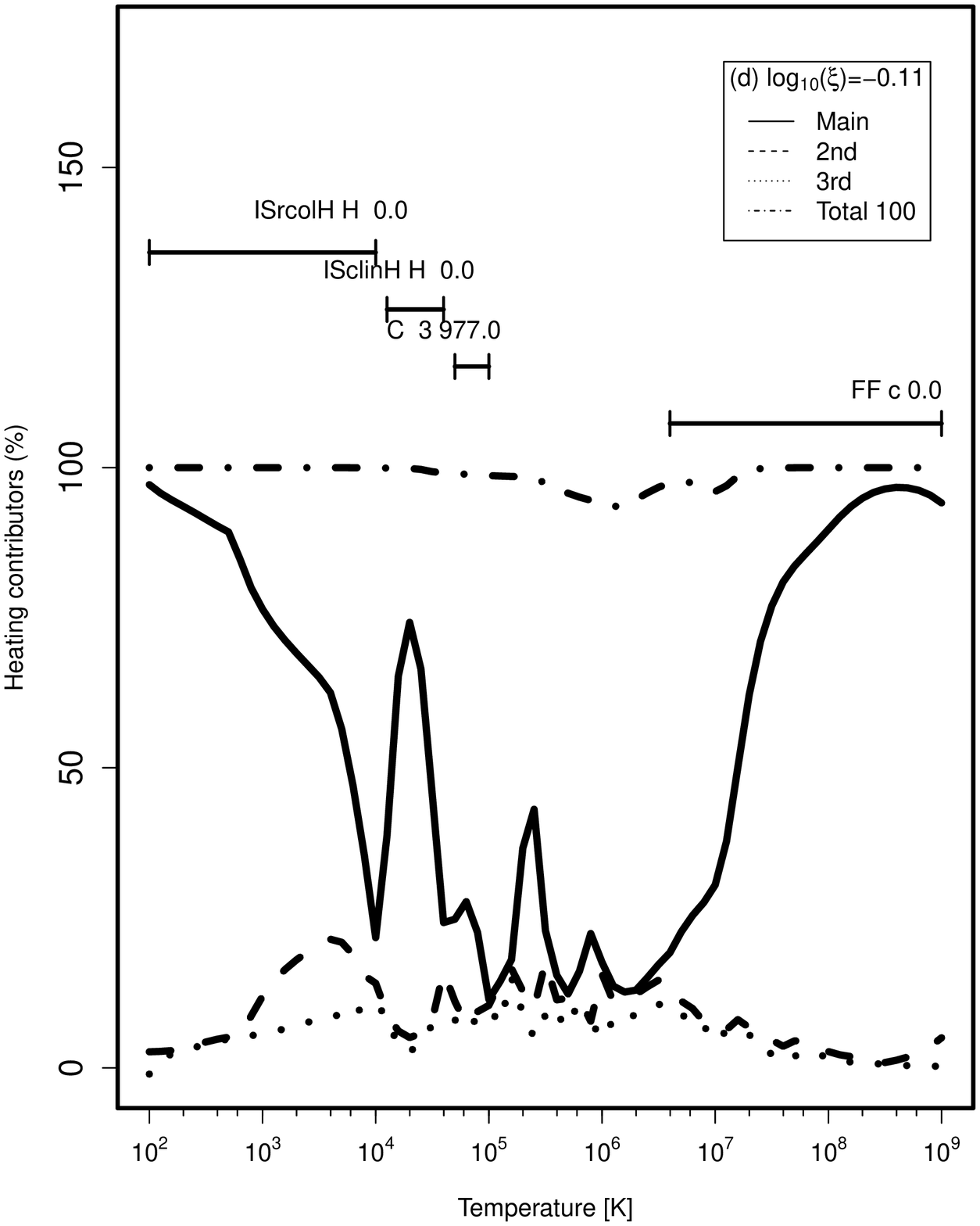}
\end{minipage} 
\caption{
Main agents contributing to the cooling rates as included in the tables.
A complex interplay between radiative and collisional recombination,
collisional de-excitation of low and high ionization species, UTA,
and free-free processes can be observed. For these plots we have used a
bremsstrahlung SED with $ T_X=1.16 \times 10^8$ K, in order to
directly see the different physical mechanisms at operation
for a gas with $n_H=10^7$ \cmd \sp at
ionization parameters $\log(\xi)$:
(a) 2.90, (b) 1.90, (c) 0.90, and (d) -0.11 [\myxi].
The main agents (solid lines)
are also labeled by straight horizontal lines to denote the temperature range
where they contribute. The second (dotted lines) and third (dashed lines)
contributors are also displayed. The dotted-dashed lines correspond to the
100 main cooling agents.
\label{coolAG1}}
\end{figure}
\begin{figure}[ht]
\rotatebox{-90}{
\begin{minipage}[b]{0.60\linewidth}
\centering
\includegraphics[width=0.95\linewidth]{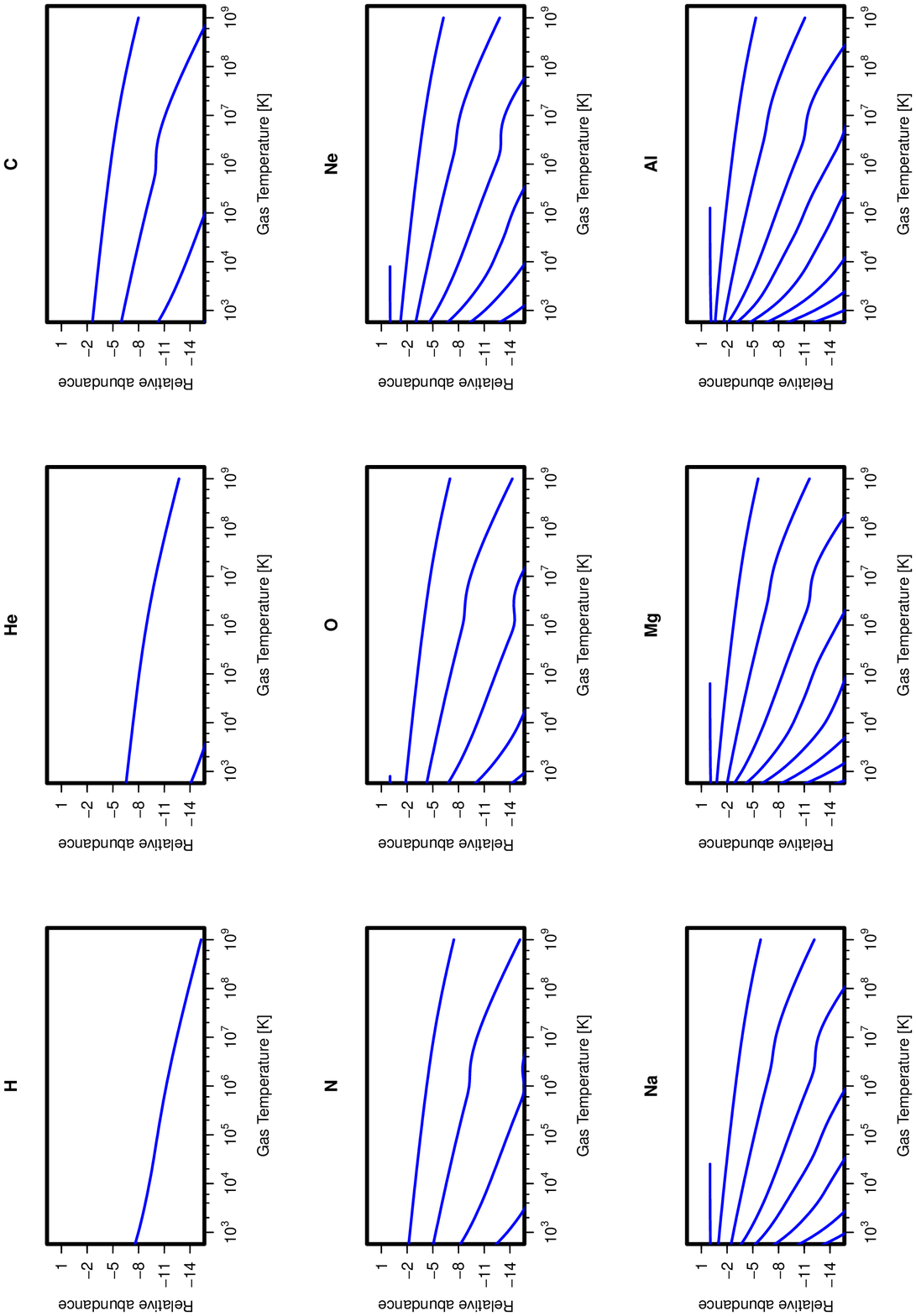}
\end{minipage} 
\begin{minipage}[b]{0.60\linewidth}
\centering
\includegraphics[width=0.95\linewidth]{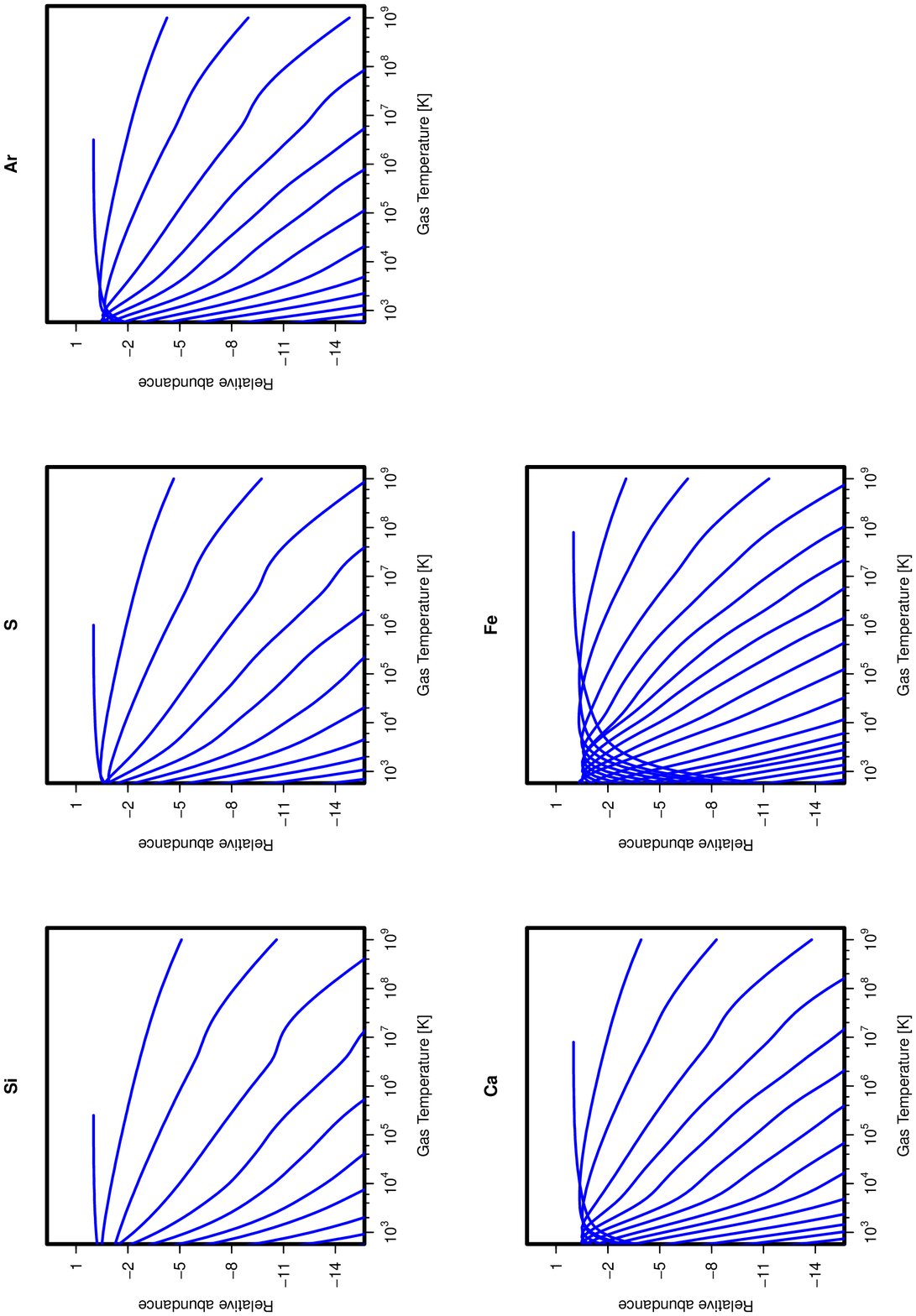} 
\end{minipage} 
}
\caption{
Relative abundances of all the elements included in our calculations
(H, He, C, N, O, Ne, Na, Mg, Al, Si, S, Ar, Ca, and Fe) as a function
of temperature. For these plots we have used $\theta=0^{\circ}$, SED1,
$f_{\rm disk}=0.8$, $f^{\rm pl}_{X}=0.2$, and $n_H=10^8$ \cmd.
\label{iofrac1}}
\end{figure}


\end{document}